\documentclass[twocolumn,showpacs,preprintnumbers,amsmath,amssymb,superscriptaddress]{revtex4-1}
\usepackage{graphicx}% Include figure files
\usepackage{dcolumn}% Align table columns on decimal point
\usepackage{bm}% bold math
\usepackage{tabularx}
\usepackage{ulem} 

\usepackage{color}   %May be necessary if you want to color links
\usepackage{hyperref}
\hypersetup{
    colorlinks=true, %set true if you want colored links
    linktoc=all,     %set to all if you want both sections and subsections linked
    linkcolor=blue,  %choose some color if you want links to stand out
}

\makeatletter
\newcommand{\colorcaption}[2][]{%
  \begingroup%
  \renewcommand{\@caption@fignum@sep}{ (Color online). }%
  \caption[#1]{#2}%
  \endgroup%
}
\makeatother
\begin{document}
%\title{Electron spectra for second-order forbidden non-unique $\beta^-$ decay of $^{24}$Na and $^{36}$Cl {\color {red} from shell model with $ab~initio$ approaches}  }

\title{ Second-forbidden nonunique $\beta^-$ decays of 
$^{24}$Na and $^{36}$Cl assessed by the nuclear shell model}
\author{Anil Kumar}
\email[]{akumar5@ph.iitr.ac.in}
\affiliation{Department of Physics, Indian Institute of Technology Roorkee,
Roorkee 247 667, India}
%\email[]{akumar5@ph.iitr.ac.in}
\author{Praveen C. Srivastava}
\email{Corresponding author: praveen.srivastava@ph.iitr.ac.in}
\affiliation{Department of Physics, Indian Institute of Technology Roorkee,
Roorkee 247 667, India}
\author{Joel Kostensalo}
\email[]{joel.j.kostensalo@student.jyu.fi}
%\affiliation{University of Jyväskylä, Department of Physics, P.O. Box 35 (YFL), FI-40014, University of Jyv\´äskylä, Finland}
\author{Jouni Suhonen}
\email[]{jouni.t.suhonen@jyu.fi}
\affiliation{University of Jyvaskyla, Department of Physics, P.O. Box 35 (YFL), FI-40014, University of Jyvaskyla, Finland}

\date{\hfill \today}
%%%%%%%%%%%%%%%%%%%%%%%%%%%%%%%%%%%%%%%%%
%\bibliographystyle{prsty}
\begin{abstract}

 We have performed a systematic study of the log$ft$ values,  shape factors and electron spectra for the  second-forbidden nonunique $\beta^-$ decays of $^{24}$Na$(4^+) \rightarrow ^{24}$Mg$(2^+)$ and
$^{36}$Cl$(2^+) \rightarrow ^{36}$Ar$(0^+)$ transitions under the framework of the nuclear shell model.
% We have also calculated log$ft$ values of these transitions. 
 We have performed  the shell model calculations in the $sd$ model space, 
 using more recent microscopic effective interactions such as Daejeon16, chiral N3LO, and JISP16. These interactions are derived 
from the no-core shell model wave functions using Okubo-Lee-Suzuki transformation.
 For comparison, we have also shown the results obtain from the phenomenological USDB interaction. 
To test the predictive power of these interactions first  we have computed 
low-lying energy spectra of parent and daughter nuclei involved in these transitions.  The computed results for energy spectra, nuclear matrix elements, log$ft$ values, shape factors, electron spectra and decomposition of the integrated shape factor are reported and compare  
with the available experimental data.   
\end{abstract}

\maketitle

\section{Introduction}\label{Introduction}
The $\beta$ decay  plays  an important role in astrophysics e.g. for the $r$ process \cite{Langanke}.  
In the nuclear chart,    there are selected candidates for double beta decays,
but on the other hand,  there are several  potential candidates known for forbidden beta decay.
 Out of these, only around 27 possible  candidates  of second-forbidden nonunique beta decay is observed  as reported in Ref.  \cite{Singh1998}.  
Recently,  a  new candidate  is observed corresponding to second-forbidden nonunique  decay of $^{20}$Fe$(2^+)\rightarrow^{20}$Ne${(0^+)}$ from ground-state-to-ground-state transition \cite{Kirsebom2019PRC,Kirsebom2019PRL,suzuki}. This study could change our understanding of the fate of intermediate-mass stars.  The comprehensive review on the theoretical and experimental status of single and double beta decay is recently reported in Ref.  \cite{Jouni_review}.

In the beta decay based on the  value of angular momentum ($l$) we can characterize any decay as allowed or forbidden. 
The $l=0$  decays are called as ``allowed" while the $l>0$  decays are called as ``forbidden". 
Further, we can divide decays  as forbidden unique (FU) and forbidden nonunique (FNU). 
In the case of FU, the total angular momentum $K=l+1$, whereas in FNU decay $K=l$.
The $\beta$ decay half-life of the 4th forbidden nonunique decay of $^{50}$V using nuclear shell model is reported in Ref. \cite{mika2014}. The 4th forbidden nonunique ground-state-to-ground-state $\beta^-$ decay branches of $^{113}$Cd and $^{115}$In using the microscopic quasiparticle-phonon model and the nuclear shell model is reported in  Refs. \cite{mika2016,mika2017}. Also in these references the half-life method \cite{mika2016} and spectrum-shape method (SSM) \cite{mika2017}  are reported to extract the value of axial-vector coupling constant $g_A$.

Studies of  the forbidden beta decay using the nuclear shell model with phenomenological interactions are available in the literature. With the recent progress in the $ab~initio$ approaches for nuclear structure study, it is highly desirable to see how these interactions are able to predict nuclear observables such as forbidden beta decay.   Recently, shell model results for allowed beta decay properties of $sd$, $fp$ and $fpg$  shell nuclei are reported by us in Refs. \cite{anil,vikas,archana,vikas2}.

In the present work, our aim is to study second-forbidden nonunique $\beta^-$ transitions of
$^{24}$Na$(4^+)\rightarrow^{24}$Mg$(2^+)$ and $^{36}$Cl$(2^+)\rightarrow^{36}$Ar$(0^+)$ 
using $ab~initio$ interactions. Beta decay transitions in these nuclei have been calculated and compared with the available experimental data to test the quality of the $ab~initio$ interaction wave functions. 
A theoretical attempt  has been made in the past to calculate the beta decay transition observable of $^{36}$Cl \cite{Sadler1993}. However, no theoretical estimate is found in the literature for the beta-decay of $^{24}$Na and also no experimental shape factors and electron spectra  are found in the literature. Thus, our theoretical  predictions for the beta decay of $^{24}$Na are useful for the future experiments. In this work, we have computed the log$ft$ values, shape factors and electron spectra of these branches. We have constrained the relativistic nuclear matrix element based on conserved vector current (CVC) theory and test the role of this matrix element in the shape factors and electron spectra. In order to test our computed wave functions, first we have computed the low-lying energy spectra of $^{24}$Na, $^{24}$Mg, $^{36}$Cl, and $^{36}$Ar and compare them with the available experimental energy spectra \cite{nndc}.

This article is organized as follows. In Sec. \ref{Formalism} we give a short overview of the theoretical formalism for the  $\beta^-$ decay and details about microscopic effective interactions. Results and discussions corresponding to low-lying energy spectra, nuclear matrix elements, log$ft$ values,  shape factors, electron spectra and decomposition of the integrated shape factors are reported in Sec.  \ref{Results}.
Finally, in Sec. \ref{Conclusions} we draw the conclusions.

%%%%%%%%%%%%%%%%%%%%%%%%%%%%%%%%%%%%%%%%%%%%%%%%%%%%%%%%%%%%%%%%%%%%%%%%%%%%%%%%%%%%%%%%%%%%%%%%%%
\section{Theoretical Formalism} \label{Formalism}
In the section (\ref{beta}), we discuss the theory of forbidden $\beta^-$ decay, and
the shape of the electron spectra.  Section (\ref{models}), give the details about
the valence space and microscopic effective interactions used in the present work.

\subsection{Beta Decay Theory:} \label{beta}

The full details of formalism for both allowed and forbidden types of the $\beta$ decay are available  in the literature by Behrens and B$\ddot{\text{u}}$hring \cite{behrens1982} (see also Ref. \cite{hfs1966}). The generalized framework of the forbidden  nonunique  $\beta$ decay theory  is available in the Refs. \cite{mst2006,ydrefors2010,mika2017}.
When the beta decay process is described as a point-like interaction vertex with an effective Fermi coupling constant $G_\text{F}$, the probability of the electron emission in the kinetic energy interval $W_e$ and $W_e+dW_e$  is expressed as 
\begin{equation} \label{eq1}
\begin{split}
P(W_e)dW_e & = \frac{G_\text{F}^2}{(\hbar{c})^6}\frac{1}{2\pi^3\hbar}C(W_e) \\
 & \times{p_ecW_e(W_0-W_e)^2F_0(Z,W_e)dW_e}.
\end{split}
\end{equation}

Where the $C(W_e)$ is the shape factor containing the nuclear structure information, and $W_0$ is
the endpoint energy of   the  $\beta$ spectrum. The factor  $F_0(Z, W_e)$ is the Fermi function, which takes into account 
Coulombic interaction between the daughter nucleus and $\beta$ particle, and Z is the proton number of the final nucleus.
Furthermore, $p_e$  and $W_e$ are the momentum and energy of the emitted electron, respectively.    

The partial half-life of the $\beta$ decay is expressed as 
\begin{eqnarray}\label{hf}
t_{1/2}=\frac{\text{ln}(2)}{\int_{m_ec^2}^{W_0}P(W_e)dW_e},
\end{eqnarray}
where $m_e$ is the mass of the electron. For the convenience, Eq. (\ref{hf}) can be expressed in the form 

\begin{eqnarray}\label{hf1}
t_{1/2}=\frac{\kappa}{\tilde{C}},
\end{eqnarray}
where $\tilde{C}$ is the unitless integrated shape factor, and the constant $\kappa$ has the value 
\begin{eqnarray}
\kappa=\frac{2\pi^3\hbar^7\text{ln(2)}}{m_e^5c^4(G_\text{F}\text{Cos}\theta_\text{C})^2}=6147s,
\end{eqnarray}
where  $\theta_C$ is the Cabibbo angle and the usual dimensionless kinematics quantities are
defined as $w_0=W_0/m_ec^2$, $w_e=W_e/m_ec^2$, and $p=p_ec/m_ec^2=\sqrt{(w_e^2-1)}$, then the dimensionless integrated shape factor  $\tilde{C}$ can be expressed as  
\begin{eqnarray} \label{tc}
\tilde{C}=\int_1^{w_0}C(w_e)pw_e(w_0-w_e)^2F_0(Z,w_e)dw_e.
\end{eqnarray}
The comparative half-life, or the $ft$ values, is obtained by multiplying the partial half-life with 
the following dimensionless integrated Fermi function
\begin{eqnarray}
f_0=\int_1^{w_0}pw_e(w_0-w_e)^2F_0(Z,w_e)dw_e,
\end{eqnarray}
but $ft$ values are usually large, so it is normally  expressed in term of ``log$ft$" values \cite{suhonen2007}. The log$ft$ values is defined as
\begin{eqnarray}
\text{log}ft=\text{log}_{10}\big({f_0t_{1/2}[s]}\big).
\end{eqnarray}
The shape factor $C(w_e)$ in Eq. (\ref{tc}) for pure Gamow-Teller transition is defined as
\begin{eqnarray}\label{mgt}
C(w_e)=\frac{g_A^2}{2J_i+1}|\mathcal{M}_{\text{GT}}|^2, 
\end{eqnarray}
where the $J_i$ is the angular momentum of the initial state, $g_A$ is the axial-vector coupling constant, and the $\mathcal{M}_\text{GT}$ is  the Gamow-Teller nuclear matrix element \cite{suhonen2007}. Which is defined as
\begin{equation} \label{eq8}
\begin{split}
\mathcal{M}_\text{GT} & \equiv(\xi_f{J_f}\|\sigma\|\xi_i{J_i}) \\
 &   =\sum_{pn}\mathcal{M}_\text{GT}(pn)(\xi_fJ_f\|[c_p^\dagger\tilde{c_n}]_1\|\xi_iJ_i),
\end{split}
\end{equation}
 where  $\mathcal{M}_\text{GT}(pn)$ is the single particle matrix elements (SPMEs). In case of forbidden nonunique beta decay the  form of the shape factor $C(w_e)$ in Eq. (\ref{tc}) is defined as

\begin{eqnarray} \label{eq2}
%\begin{split}
C(w_e)  = \sum_{k_e,k_\nu,K}\lambda_{k_e} \Big[M_K(k_e,k_\nu)^2+m_K(k_e,k_\nu)^2 \nonumber\\
    -\frac{2\gamma_{k_e}}{k_ew_e}M_K(k_e,k_\nu)m_K(k_e,k_\nu)\Big],
%\end{split}
\end{eqnarray}

where the indices $k_e$ and $k_\nu$ ($k_e,k_\nu$=1,2,3,...) are positive integers, which are emerging from the partial-wave expansion of the lepton wave functions and $K$ is the order of forbiddeness of the transition. The  nuclear structure information is contained in the quantities $M_K(k_e,k_\nu)$ and $m_K(k_e,k_\nu)$, which are complicated combinations of different nuclear matrix elements (NMEs) and leptonic phase-space factors. The factor $\lambda_{k_e}$ is the Coulomb function and expressed  as
\begin{equation}
\lambda_{k_e}=\frac{F_{k_e-1}(Z,w_e)}{F_0(Z,w_e)},
\end{equation}

where $F_{k_e-1}(Z,w_e)$ is the generalized Fermi function \cite{mst2006,mika2017}, which is expressed as
\begin{eqnarray} \label{eq3}
F_{k_e-1}(Z,w_e) &=4^{k_e-1}(2k_e)(k_e+\gamma_{k_e})[(2k_e-1)!!]^2e^{\pi{y}} \nonumber \\
 & \times\left(\frac{2p_eR}{\hbar}\right)^{2(\gamma_{k_e}-k_e)}\left(\frac{|\Gamma(\gamma_{k_e}+iy)|}{\Gamma(1+2\gamma_{k_e})}\right)^2.
\end{eqnarray}

The auxiliary  quantities are defined as $\gamma_{k_e}=[k_e^2-(\alpha{Z})^2]^{1/2}$ and $y=(\alpha{Zw_e}/p_ec)$, where $\alpha=1/137$ is the fine structure constant. 

The  nuclear matrix elements (NMEs) are given by
\begin{eqnarray} \label{nme}
^{V/A}\mathcal{M}_{KLS}^{(N)}(pn)(k_e,m,n,\rho)=\frac{1}{\sqrt{2J_i+1}} \nonumber\\
 \times\sum_{pn}{^{V/A}m_{KLS}^{(N)}}(pn)(k_e,m,n,\rho)(\psi_f\|[c_p^\dagger{\tilde{c}_n}]\|\psi_i).
\end{eqnarray}

The nuclear matrix elements are divided in two  parts: first part $^{V/A}m_{KLS}^{(N)}(pn)(k_e,m,n,\rho)$ is called the 
single-particle matrix element  and second part $(\psi_f\|[c_p^\dagger{\tilde{c}_n}]\|\psi_i)$ is the reduced one-body 
transition density (OBTD) between the initial ($i$) and final ($f$) nuclear states. The   single-particle matrix
elements  characterizes the properties of the transition operators, so they are the same for all 
nuclear models. But the OBTDs are the nuclear model dependent. In the present work the SPMEs are calculated 
using harmonic-oscillator wave functions (see  Refs. \cite{mst2006,mika2017}). The summation of Eq. (\ref{nme}) runs over 
the proton ($p$) and neutron ($n$) single-particle states.

The shape factor $C(w_e)$ (\ref{eq2}) can be decomposed into vector, axial-vector,
and mixed vector-axial-vector components \cite{mika2017,mika2016,joel12017,joel22017,jouni2017} in the form
\begin{eqnarray}\label{dcmp}
C(w_e)=g_V^2C_V(w_e)+g_A^2C_A(w_e)+g_Vg_AC_{VA}(w_e) {\color{red} .}
\end{eqnarray}
After the integration of Eq.(\ref{dcmp}) with respect to electron kinetic energy, we get the analogous expression to Eq. (\ref{tc}) 
for the integrated shape function $\tilde{C}$ 
\begin{eqnarray}\label{intc}
\tilde{C}=g_V^2\tilde{C}_V+g_A^2\tilde{C}_A+g_Vg_A\tilde{C}_{VA}.
\end{eqnarray}

In Eq. (\ref{dcmp}) the shape factors $C_i$ are functions of the electron kinetic energy, while the integrated shape factors $\tilde{C_i}$ in Eq. (\ref{intc})  are just constant numbers.

\begin{figure*}
\centering
\includegraphics[width=8.7cm,height=8.0cm,clip]{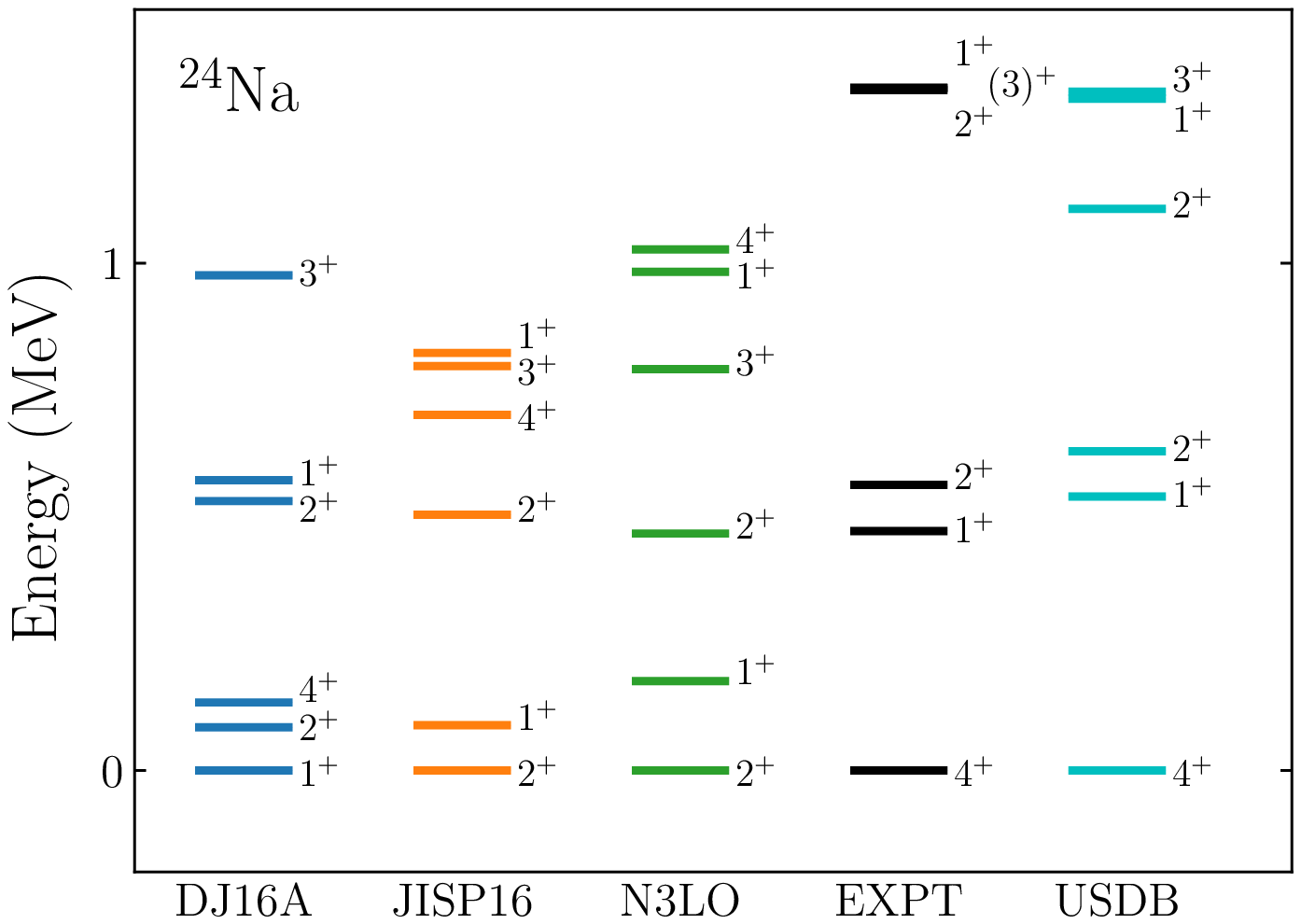}
\includegraphics[width=8.7cm,height=8.0cm,clip]{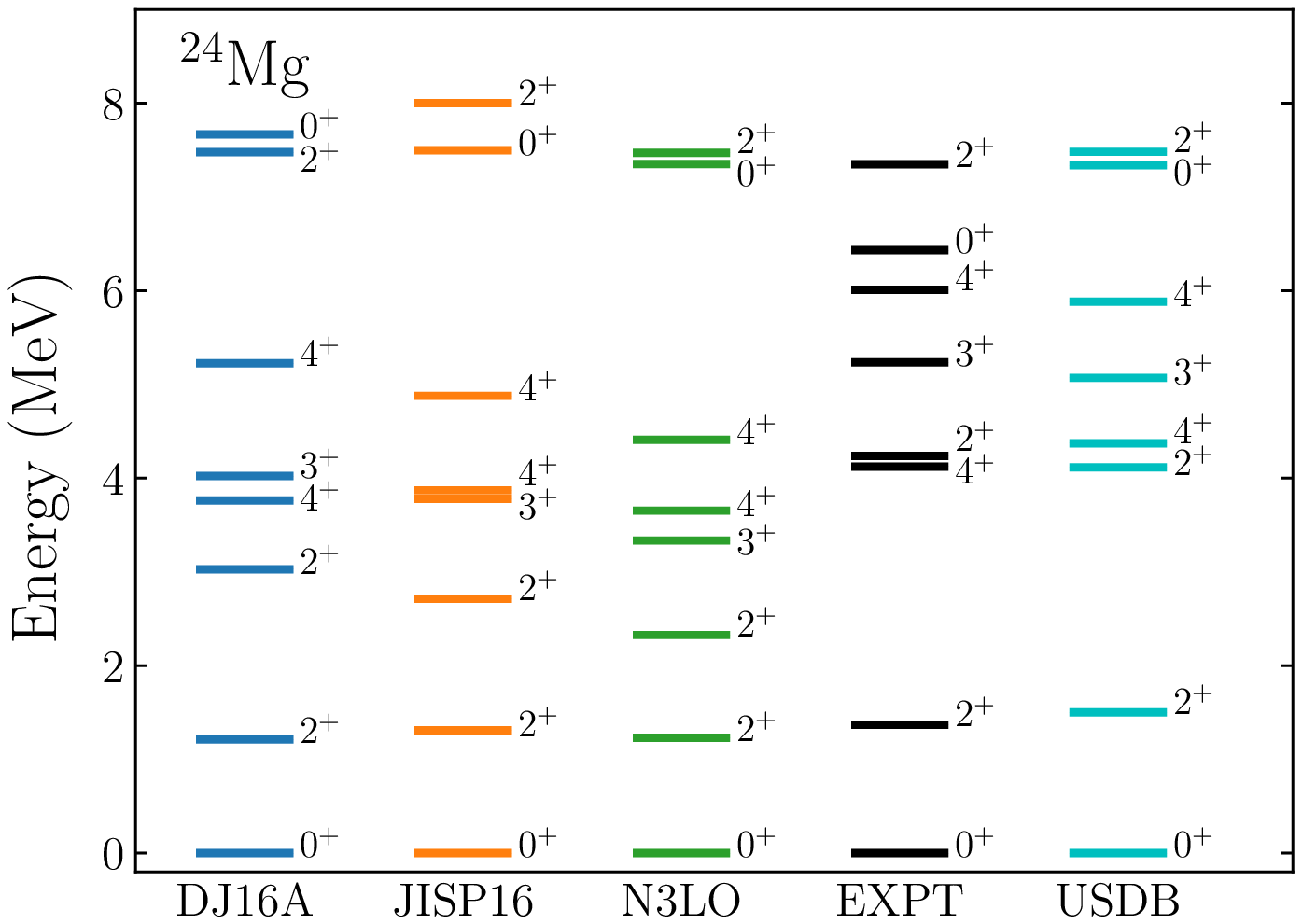}
 \caption{\label{Enrgna_fig} Comparison of calculated and experimental \cite{nndc}  low-lying energy spectra for positive parity 
 states of $^{24}$Na and $^{24}$Mg  from  microscopic and USDB interactions.}
\end{figure*}

\begin{figure*}
\centering
\includegraphics[width=8.7cm,height=8.0cm,clip]{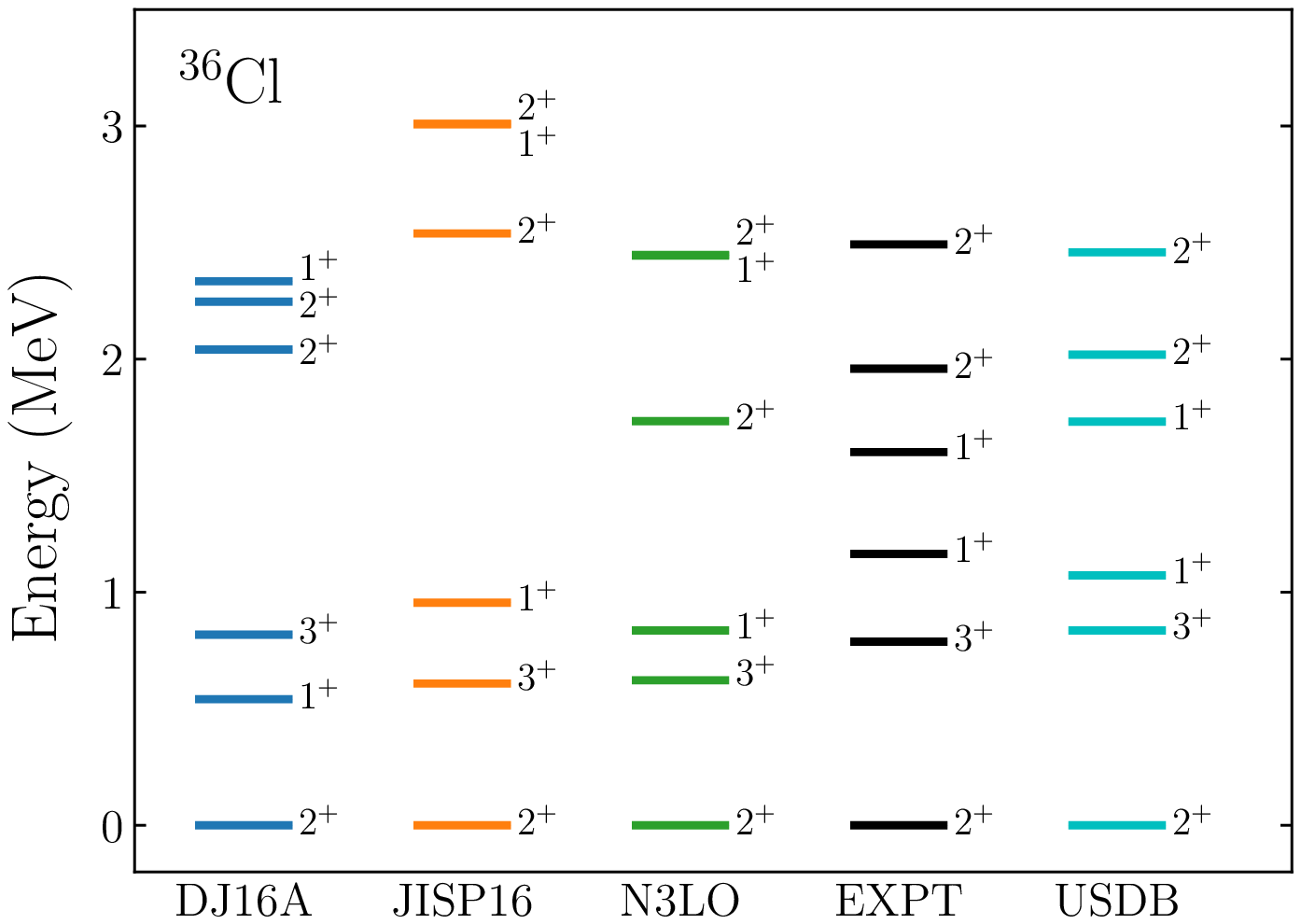}
\includegraphics[width=8.7cm,height=8.0cm,clip]{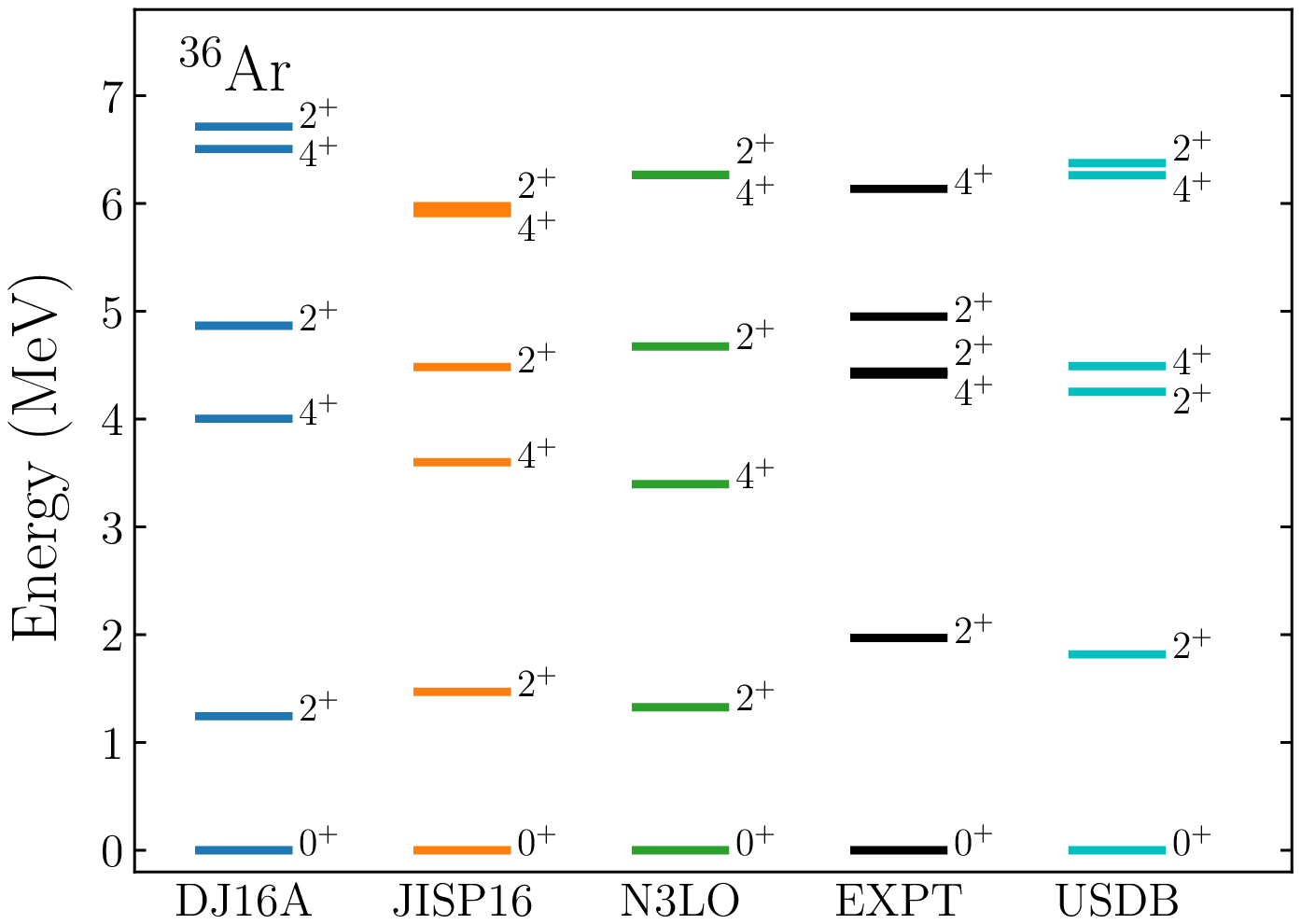}
 \caption{\label{Enrgcl_fig} Comparison of calculated and experimental \cite{nndc}  low-lying energy spectra for positive parity 
 states of $^{36}$Cl and $^{36}$Ar  from  microscopic  and USDB interactions. }
\end{figure*}

%\begin{figure*}
%\centering
%\includegraphics[width=\columnwidth]{Energy_24Na.eps}
%\includegraphics[width=\columnwidth]{Energy_24Mg.eps}
% \caption{\label{Enrgna_fig} Comparison of calculated and experimental \cite{nndc}  low-lying energy spectra %for positive parity 
% states of $^{24}$Na and $^{24}$Mg  from  microscopic and USDB interactions.}
%\end{figure*}

%\begin{figure*}
%\centering
%\includegraphics[width=\columnwidth]{Energy_36Cl.eps}
%\includegraphics[width=\columnwidth]{Energy_36Ar.eps}
% \caption{\label{Enrgcl_fig} Comparison of calculated and experimental \cite{nndc}  low-lying energy spectra %for positive parity 
% states of $^{36}$Cl and $^{36}$Ar  from  microscopic  and USDB interactions. }
%\end{figure*}

%%%%%%%%%%%%%%%%%%%%%%%%%%%%%%%%%%%%%%%%%%%%%%%%%%%%%%%%%%%%%%%%%%%%%%

\subsection{ADOPTED MODEL SPACE AND  HAMILTONIANS}\label{models}

 In the present work shell model  calculations for the low-lying energy spectra,  log$ft$ values, shape factors and electron spectra of the $\beta^-$ decay branches of $^{24}$Na and $^{36}$Cl were 
performed in the $sd$ model space.  In this framework we have calculated the OBTDs related to the NMEs of the shape factor.  For the $sd$ model space, we have used the three microscopic effective interactions: DJ16A \cite{dj16a2019}, JISP16 \cite{Dikmen2015}, N3LO \cite{Dikmen2015}. These interactions are obtained from the no-core shell model (NCSM) wave functions via  the Okubo-Lee-Suzuki (OLS) unitary transformation \cite{okubo1954,suzuki1980,suzuki1982}.   We have also compared our results with the  phenomenological USDB effective interaction \cite{usdb2006}. The interaction ``DJ16" \cite{dj16a2019}  is obtained from the Daejeon16 $NN$ potential \cite{dj162016}. After the monopole modification of ``DJ16", interaction is labeled as ``DJ16A" interaction \cite{dj16a2019}. In this work, we have used  DJ16A interaction for further calculations.
The OBTDs for NMEs were computed by the shell-model code NuShellX \cite{brown2014}.   
For the evaluation of the many-body matrix elements, we have used the single-particle matrix elements expression given in Ref. \cite{behrens1982}. In our shell-model calculations, we have used the single-particle matrix elements in the Condon-Shortley \cite{Condon1951} phase convention.
%%instead of Biedenharn-Rose \cite{Biedenharn1953} phase convention. 

%%%%%%%%%%%%%%%%%%%%%%%%%%%%%%%%%%%%%%%%%%%%%%%%%%%%%%%%%%%%%%%%%%%%%%%%%
\section{Results and Discussions}\label{Results}

  In this section we present our calculated results of low-lying energy spectra, nuclear matrix elements, logft values, shape factors, electron spectra and decomposition of the integrated shape factors
for the second-forbidden nonunique $\beta^-$ transitions of $^{24}$Na$(4^+) \rightarrow ^{24}$Mg$(2^+)$ and
$^{36}$Cl$(2^+) \rightarrow ^{36}$Ar$(0^+)$.

Previously, the log$ft$ values and shape factors of the second-forbidden beta decay of 
$^{36}$Cl \cite{Sadler1993} have been reported by applying two different nuclear models: with the pure  $1d_{3/2}\rightarrow1d_{3/2}$ transitions and shell-model with $sd$ shell configuration space.

Recently, much progress has been achieved in developing modern effective interactions
for the shell model calculations. Thus we have revisited  
calculation for $^{36}$Cl and also the first time   for  $^{24}$Na with recently developed microscopic  (DJ16A, N3LO, and JISP16) and phenomenological (USDB) interactions in the $sd$ model space. Our results for $^{24}$Na will be useful when compared with upcoming experimental data.

 Below we have presented low-lying energy spectra ( Figs. \ref{Enrgna_fig}-\ref{Enrgcl_fig}), nuclear matrix elements (Table \ref{mgt_tab} and \ref{fnunmes_tab}), log$ft$ values (Table \ref{logft_gt} and \ref{logft_tab}), shape factors and electron spectra (Figs. \ref{36cl_shape}-\ref{24na_shape}).
The low-lying energy spectra are discussed in Sec. \ref{energy_spec}. The $\beta$ decay nuclear matrix elements 
and log$ft$ values are discussed in Sec. \ref{nmes_sec}. Results of  the  shape factors and electron spectra are presented in  Sec. \ref{ele_spec}.  Decomposition of the integrated shape factor are discussed in Sec. \ref{shape_d}.

\subsection{Low-lying energy spectra} \label{energy_spec}

In Fig. \ref{Enrgna_fig}, we show the low-lying energy spectra of $^{24}$Na and $^{24}$Mg. In the case of $^{24}$Na, 
the ground state  (g.s.) $4^+$ is correctly reproduced by USDB interaction, while the other microscopic effective interactions
N3LO, and JISP16 give the $2^+$  as a g.s., and DJ16A  predict g.s. as $1^+$.
The low-energy spectrum 
of the well known $sd$-shell rotor nucleus $^{24}$Mg is already shown in Ref. \cite{dj16a2019} for all the interactions 
that we  have used in the present work.  For $^{24}$Mg,  the  $0_{g.s.}^+$ and $2_1^+$ are relatively well
described by all the interactions. The computed  $2_1^+$ state is obtained at 1.213, 1.310, 1.231, and 1.502 MeV 
corresponding to DJ16A, JISP16, N3LO, and USDB, respectively, while the corresponding experimental
value is 1.369 MeV. The theoretical low-lying energy spectra of $^{36}$Cl, and $^{36}$Ar are shown in Fig. \ref{Enrgcl_fig}
in comparison with the experimental data. The g.s. is correctly reproduced by the microscopic 
(DJ16A, JISP16, and N3LO) and USDB interactions for $^{36}$Cl and $^{36}$Ar.  For $^{36}$Cl, the order of  $3_1^+$ and $1_1^+$ states 
are  correctly reproduced from the JISP16, N3LO, and USDB interactions as 
in the experimental data, while the DJ16A interaction invert the order of these states. In the case of $^{36}$Ar, 
the calculated  $2_1^+$  state from the DJ16A, JISP16, N3LO, and USDB interactions are close to the
experimental data. So, in general, the comparison of the computed low-lying energy levels are in good agreement with the  experimental
data for $^{24}$Na, $^{36}$Cl, and $^{36}$Ar. In the present work we have taken  $Q$ values from the 
experimental data \cite{nndc} for further calculations listed in the Table~\ref{logft_gt} and \ref{logft_tab}.

\subsection{  Nuclear matrix elements  and log$ft$ values} \label{nmes_sec}

The nuclear matrix elements contain the nuclear-structure information. 
The Gamow-Teller  matrix elements $\mathcal{M}_{\text{GT}}$ calculated from the  microscopic and USDB interactions 
for the allowed $\beta^-$ decays of $^{24}$Na($4^+$)$ \rightarrow ^{24}$Mg($3_1^+,4_1^+$) transitions   are presented in the Table \ref{mgt_tab} with comparison  to the   experimental data.  The experimental $\mathcal{M}_\text{GT}$ value is obtained from the  log$ft$ \cite{nndc} values corresponding  to the axial-vector  coupling constant  $g_A=1.00$. 
In the present work, we have calculated these matrix elements by using OBTDs corresponding to all microscopic  and 
USDB interactions. After that, we compare the calculated $\mathcal{M}_\text{GT}$ with the experimental data. 
For the both allowed transitions, the calculated $\mathcal{M}_\text{GT}$ values from USDB are   close
to the experimental data as compared to the microscopic interactions. In the case of  $^{24}$Na($4^+$)$ \rightarrow ^{24}$Mg($4_1^+$)  transition,
our calculated value of  $\mathcal{M}_\text{GT}$ (0.0441) from the DJ16A  is very small in comparison with the experimental data. 

 The calculated log$ft$ values of allowed  $\beta^-$ decays of $^{24}$Na($4^+$)$ \rightarrow ^{24}$Mg($3_1^+,4_1^+$) transitions  are presented in Table \ref{logft_gt} in comparison 
to the experimental data.
For the    calculation, we have used  the axial-vector coupling constant  $g_A=1.00$ and $g_A=1.27$. 
%The log$ft$ values for these transitions were calculated corresponding to the value of weak coupling constants $g_A=1.00$ and $g_A=1.27$.} 
   For the transition $4^+\rightarrow{3_1^+}$, the calculated log$ft$ 
values for $g_A=1.00$ are in nice agreement with the experimental values corresponding to USDB, also all other microscopic effective interactions are in  a reasonable agreement.
%% are fairly agreement corresponding $g_A=1.00$. 
 However, in the case of
$4^+\rightarrow{4_1^+}$   transition, the calculated log$ft$ value from DJ16A  is  larger in comparison with the experimental 
data,  but from other interactions they are close to the experimental data with
both $g_A$ values.

\begin{table}[!hb]
\vspace{-1ex}
\caption{\label{mgt_tab} Calculated Gamow-Teller  matrix elements  of the allowed $\beta^- $  
 decays from the g.s. ( $4^+$)  of $^{24}$Na to  the excited states in $^{24}$Mg  from   microscopic and USDB effective interactions.}
\begin{ruledtabular}
\begin{tabular}{crrrrr}
 & \multicolumn{5}{c}{$|\mathcal{M_\text{GT}}|$}  \\
\cline{2-6}
Transitions  &  USDB &   DJ16A & N3LO & JISP16 &  Expt \\
\hline
$4^+\rightarrow{3_1^+}$    &  0.1859  &    0.1982   & 0.2274    & 0.2108  &   0.1179\\
$4^+\rightarrow{4_1^+}$    &  0.2663  &    0.0441   & 0.1069    & 0.0839  &   0.2072 \\
\end{tabular}
\end{ruledtabular}
\end{table}

\begin{table*}[!ht]
\vspace{-1ex}
\caption{\label{logft_gt} Calculated log$ft$ values of the allowed $\beta^- $ decays from g.s. ( $4^+$)  of $^{24}$Na to  the
excited states in $^{24}$Mg from the microscopic and USDB effective interactions. }
\begin{ruledtabular}
\begin{tabular}{cccccccccccc}
   & &  \multicolumn{6}{c}{log$ft(g_A=1.00)$} &\multicolumn{2}{c}{log$ft(g_A=1.27)$} \\
\cline{4-7}
\cline{8-11}
Transitions & Q(MeV) & BR($\%$) & USDB &   DJ16A & N3LO  & JISP16 & USDB & DJ16A & N3LO & JISP16 &Expt  \\
\hline
$4^+\rightarrow{3_1^+}$ &  0.280 & 0.076  & 6.205 & 6.149 & 6.029 & 6.095 & 5.997  & 5.941  & 5.822 &  5.888  &  6.60(2)\\
$4^+\rightarrow{4_1^+}$ & 1.392 & 99.855  & 5.892 & 7.454 & 6.685 & 6.896 & 5.685  & 7.247  & 6.478 &  6.688  &   6.11(1)\\
\end{tabular}
\end{ruledtabular}
\end{table*}

\begin{table*} [!ht]
\leavevmode
%\begin{small}
\centering
\caption{\label{fnunmes_tab} Calculated leading-order nuclear matrix elements (NMEs) of 
the second-forbidden nonunique $\beta^- $  decays of $^{24}$Na and $^{36}$Cl are from  microscopic and USDB interactions.
The Coulomb-corrected NMEs are indicated by ($k_e,m,n,\rho$), when such elements exist. }
\begin{ruledtabular}
\begin{tabular}{lcccc}\\ 
Nuclear & \multicolumn{4}{c}{$^{24}$Na($4^+$)$\rightarrow${$^{24}$Mg($2^+$)}}\\
\cline{2-5} 
 
 Matrix Elements & USDB & DJ16A  & N3LO  & JISP16 \\
\hline 
%{$^{24}$Na($4^+$)$\rightarrow${$^{24}$Mg($2^+$)}}\\
$^V\mathcal{M}^{(0)}_{211}$(CVC)       & 0.023790$\pm$0.0001 & -0.018446$\pm$0.0002 & -0.020217$\pm$0.0001  &   -0.019636$\pm$0.0001   \\
$^V\mathcal{M}^{(0)}_{220}$            & 0.431273 & -0.131891   & -0.237936 & -0.187614      \\
$^V\mathcal{M}^{(0)}_{220}(1,1,1,1)$   & 0.530979 & -0.123441   & -0.264185 & -0.203108      \\
$^V\mathcal{M}^{(0)}_{220}(2,1,1,1)$   & 0.509588 & -0.110404   & -0.247587 & -0.189152      \\ 
$^A\mathcal{M}^{(0)}_{221}$            &-0.430287 & -0.482638   & -0.219655 & -0.294803      \\
$^A\mathcal{M}^{(0)}_{221}(1,1,1,1)$   &-0.524687 & -0.577264   & -0.287289 & -0.370261      \\
$^A\mathcal{M}^{(0)}_{221}(2,1,1,1)$   &-0.502493 & -0.550486   & -0.279212 & -0.356859      \\
$^A\mathcal{M}^{(0)}_{321}$            &-1.459626 & -0.758772   & -0.067127 & -0.050213      \\

\hline\\
Nuclear & \multicolumn{4}{c}{$^{36}$Cl($2^+$)$\rightarrow${$^{36}$Ar($0^+$)}}\\
\cline{2-5} 
 
 Matrix Elements & USDB & DJ16A  & N3LO  & JISP16 \\
\hline 
%{$^{24}$Na($4^+$)$\rightarrow${$^{24}$Mg($2^+$)}}\\
$^V\mathcal{M}^{(0)}_{211}$(CVC)       &-0.029375$\pm$0.0005 & -0.015943$\pm$0.0010 & -0.022497$\pm$0.0008  &   -0.007451$\pm$0.0009   \\
$^V\mathcal{M}^{(0)}_{220}$            & -5.892542 & -3.483430  & -4.705624 & -5.057782      \\
$^V\mathcal{M}^{(0)}_{220}(1,1,1,1)$   & -7.250832 & -4.357072  & -5.796787 & -6.225284      \\
$^V\mathcal{M}^{(0)}_{220}(2,1,1,1)$   & -6.955989 & -4.195245  & -5.562475 & -5.972497      \\ 
$^A\mathcal{M}^{(0)}_{221}$            & -1.249043 & -2.025348  & -1.716437 & -1.644994     \\
$^A\mathcal{M}^{(0)}_{221}(1,1,1,1)$   & -1.496326 & -2.412741  & -2.062063 & -1.979877      \\
$^A\mathcal{M}^{(0)}_{221}(2,1,1,1)$   & -1.426626 & -2.297321  & -1.967308 & -1.889710      \\

\end{tabular}
\end{ruledtabular}
%\end{small}
\end{table*}

\begin{table*}[!ht]
%\begin{small}
\centering
\caption{\label{logft_tab} Calculated log$ft$ values of the second-forbidden nonunique $\beta^-$ decays of $^{24}$Na and 
$^{36}$Cl from shell model and after constrained  the matrix element $^V\mathcal{M}^{(0)}_{211}$ 
from experimental data. For the log$ft$ calculations we have used the value of coupling constants $g_V$=1.00 and $g_A$=1.27. 
The experimental data have been taken from \cite{nndc}.}
\label{tab:ST-pn}\vspace{1ex}
\begin{ruledtabular}
\begin{tabular}{lccrrrrrr}\\

 & & & & &\multicolumn{3}{c}{log$ft$(SM)} \\
\cline{5-9}
 Transitions & Type &  Q(MeV) & BR($\%$)& USDB &  DJ16A & N3LO &  JISP16 &  Expt \\
\hline
$^{24}$Na($4^+$)$\rightarrow${$^{24}$Mg($2^+$)}   & 2nd non-unique forbidden   & 4.147 & 0.064  & 12.237 & 12.881 & 14.227 & 13.958 & 11.340(4) \\
$^{36}$Cl($2^+$)$\rightarrow${$^{36}$Ar($0^+$)}   & 2nd non-unique forbidden   & 0.710 & 98.1   & 12.635 & 13.978 & 13.120 & 12.976 & 13.321(3) \\

\hline \\
 & & & & &\multicolumn{3}{c}{log$ft$(SM+CVC)} \\
\cline{5-9}
 Transitions & Type &  Q(MeV) & BR($\%$)& USDB &  DJ16A & N3LO &  JISP16 &  Expt \\
\hline
$^{24}$Na($4^+$)$\rightarrow${$^{24}$Mg($2^+$)}   & 2nd non-unique forbidden   & 4.147 & 0.064  & 11.367 & 11.331 & 11.346 & 11.342 & 11.340(4) \\
$^{36}$Cl($2^+$)$\rightarrow${$^{36}$Ar($0^+$)}   & 2nd non-unique forbidden   & 0.710 & 98.1   & 13.221 & 13.108 & 13.153 & 13.555 & 13.321(3) \\

\end{tabular}
\end{ruledtabular}
%\end{small}
\end{table*}

For the second-forbidden nonunique $\beta^-$ decays of $^{24}$Na($4^+$)$\rightarrow^{24}$Mg(2$^+$)
and $^{36}$Cl($2^+$)$\rightarrow^{36}$Ar($0^+$), the computed NMEs  from different microscopic 
and USDB effective interactions  are presented in Table \ref{fnunmes_tab}. The relativistic matrix element $^{V}\mathcal{M}_{211}^{(0)}$ is becoming
identically zero due to limitation of our $0\hbar{\omega}$  $sd$-shell calculations for harmonic-oscillator wave functions. 
To get the value of  $^{V}\mathcal{M}_{211}^{(0)}$ matrix element non-zero  we need to perform shell model calculations in  the
 multi-$\hbar\omega$ excitations. 
However, here we follow a different approach to
calculate the $^{V}\mathcal{M}_{211}^{(0)}$ matrix element. 
We have used a approach based on CVC theory, since we have an experimental partial half-life, so we keep
the value of coupling   constants $g_V=g_A=1.0$ and try to reproduce the value of the experimental partial half-life by varying 
the matrix element $^{V}\mathcal{M}_{211}^{(0)}$. The $^{V}\mathcal{M}_{211}^{(0)}$ matrix element obtained with this approach is labeled
as ``$^{V}\mathcal{M}_{211}^{(0)}$(CVC)'' in Table \ref{fnunmes_tab}. 

The axial-vector matrix elements $^A\mathcal{M}^{(0)}_{221}$, $^A\mathcal{M}^{(0)}_{221}(1,1,1,1)$, $^A\mathcal{M}^{(0)}_{221}(2,1,1,1)$, and $^A\mathcal{M}^{(0)}_{321}$ could be affected by the quenching of axial-vector coupling constant $g_A$.
The affected value of the Gamow-Teller transition matrix element by the quenching of axial coupling constant was observed in \cite{ Towner1987}. 
In the  recent study of  the second-forbidden nonunique beta decay of $^{20}$F, 
the effect of the quenching of axial-vector coupling constant in  axial-vector matrix elements  is reported in Refs.
\cite{Kirsebom2019PRC, Kirsebom2019PRL}.
Here, we will use the value of the axial-vector coupling constant for
the two different cases, either the bare value of $g_A=1.27$ or the   quenched value of $g_A=1.00$.

In the Table \ref{logft_tab}, we presented the log$ft$ values for the  second-forbidden nonunique $\beta^-$ decays of $^{24}$Na and $^{36}$Cl calculated 
with different  microscopic  and phenomenological
interactions in  comparison with  the experimental data, and the value of coupling constants are taken  as  $g_A=1.27$ and $g_V=1.00$
for the calculations. The results with pure shell-model labeled as ``SM'', and those constrained by experimental information, 
labeled ``SM + CVC''. The prediction of log$ft$ values with SM is far from the experimental data. However, 
the agreement between the calculation with ``SM+CVC'' and the experimental value came out to be very satisfactory.

\begin{figure*}
%[!ht]
\centering
\includegraphics[width=0.90\columnwidth]{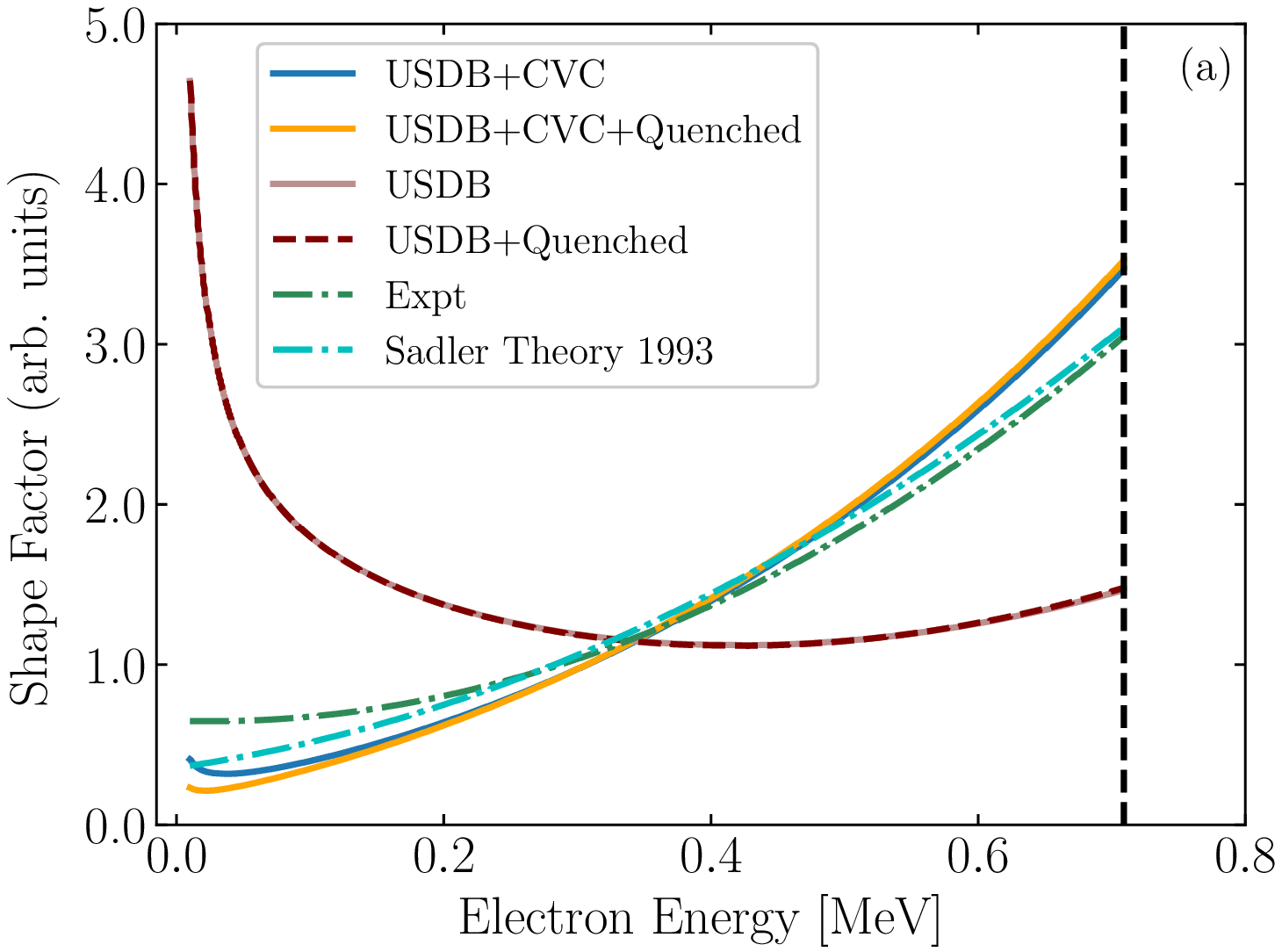}
\includegraphics[width=0.90\columnwidth]{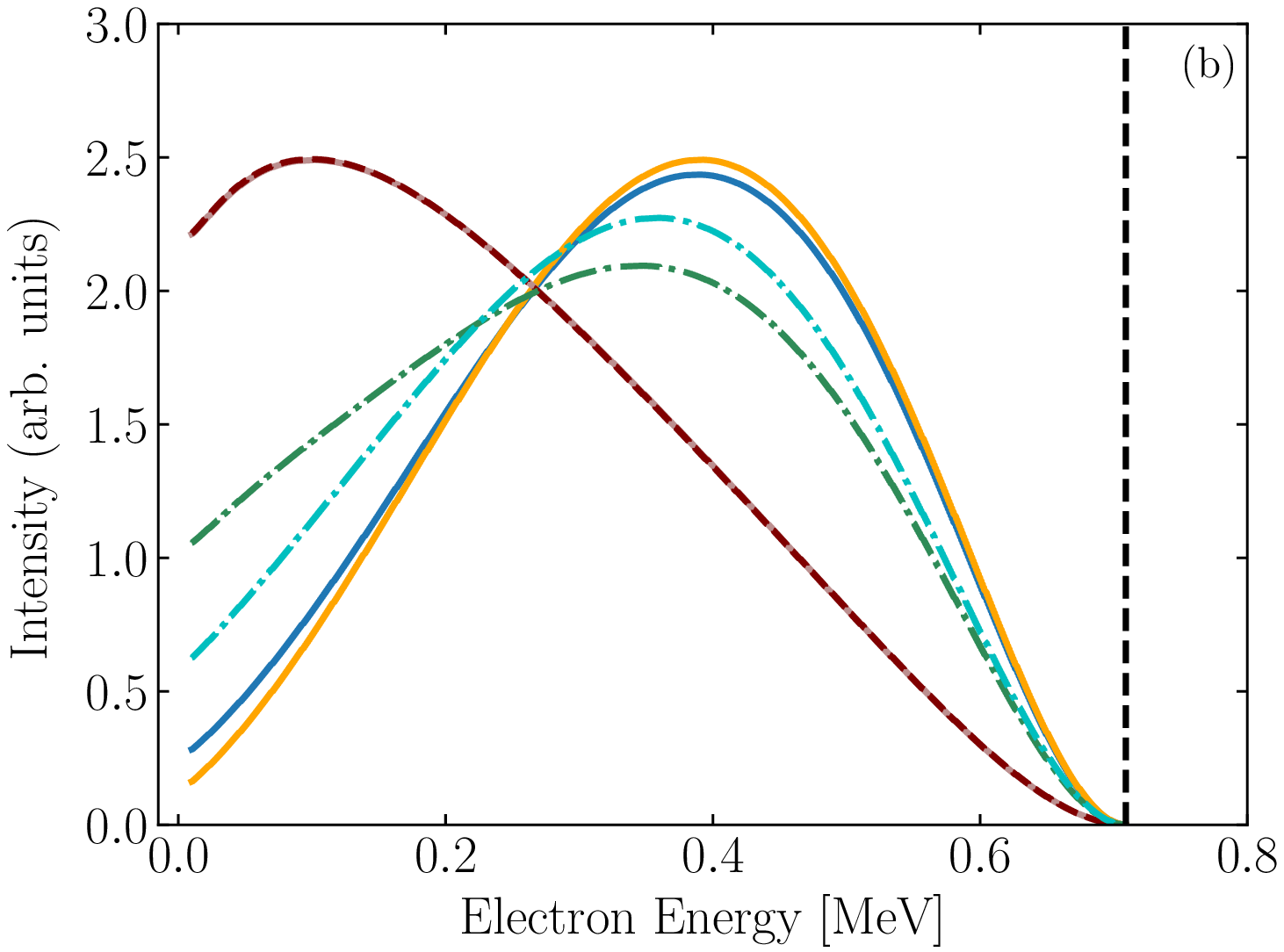}
\includegraphics[width=0.90\columnwidth]{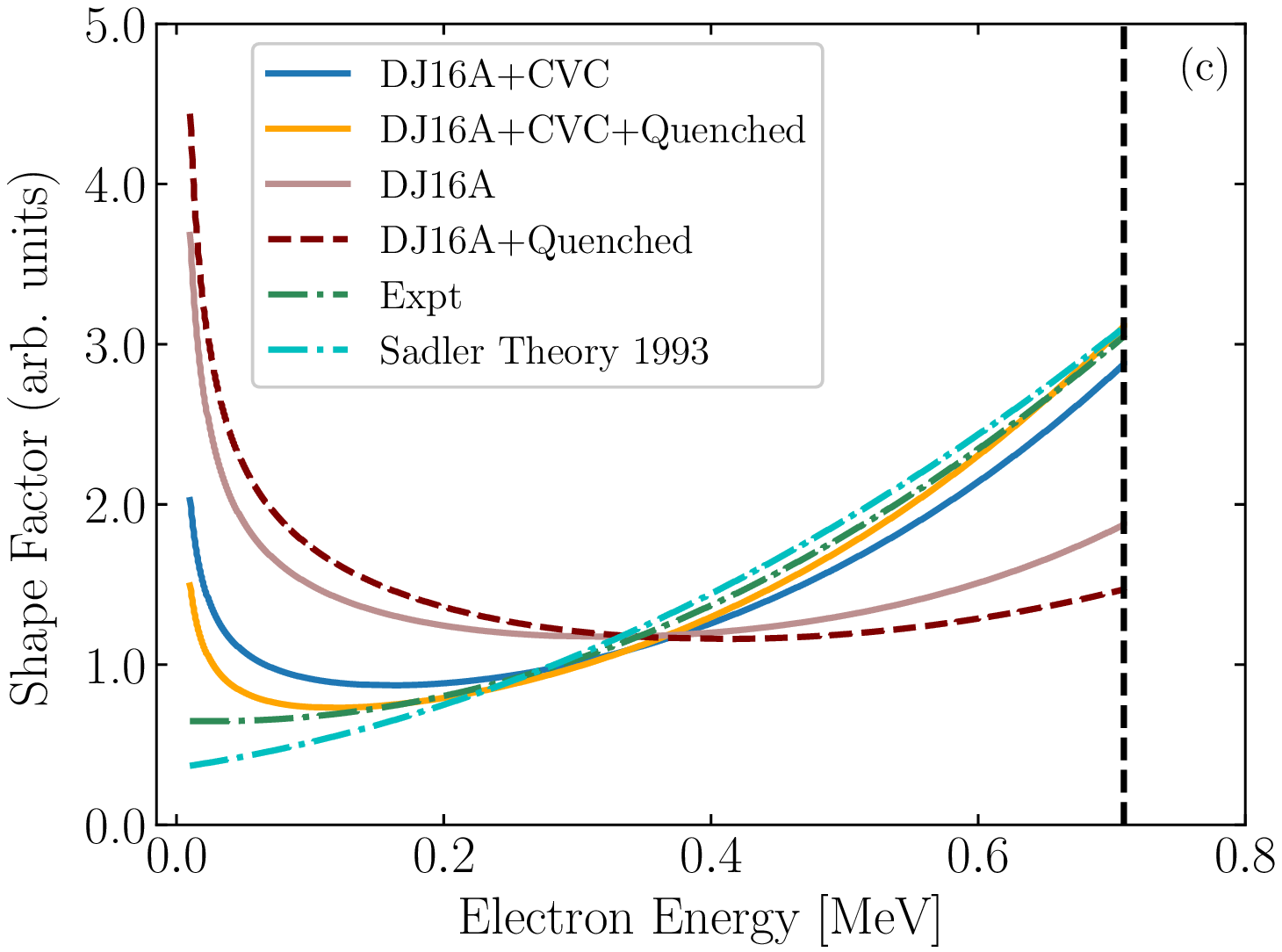}
\includegraphics[width=0.90\columnwidth]{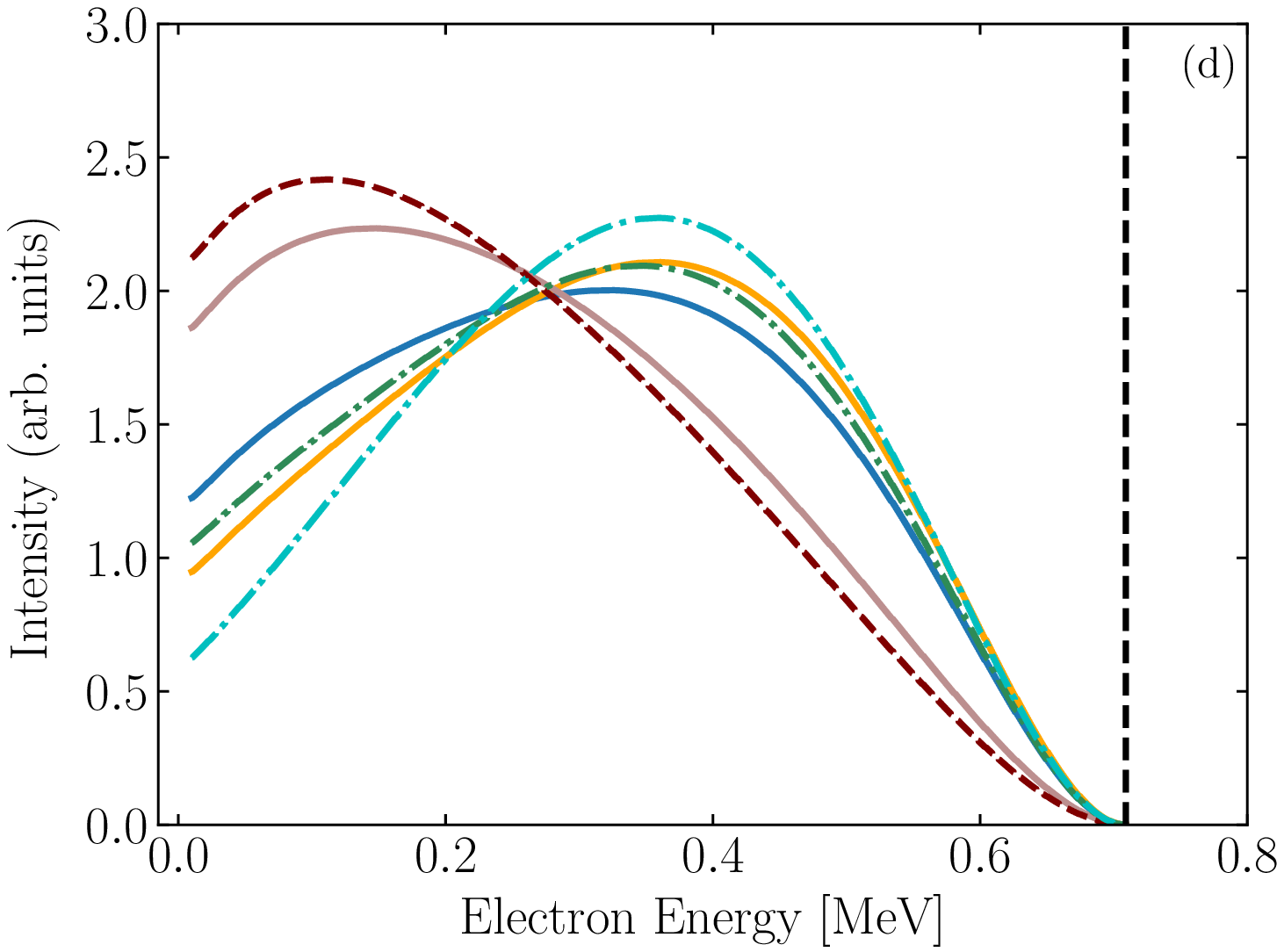}
\includegraphics[width=0.90\columnwidth]{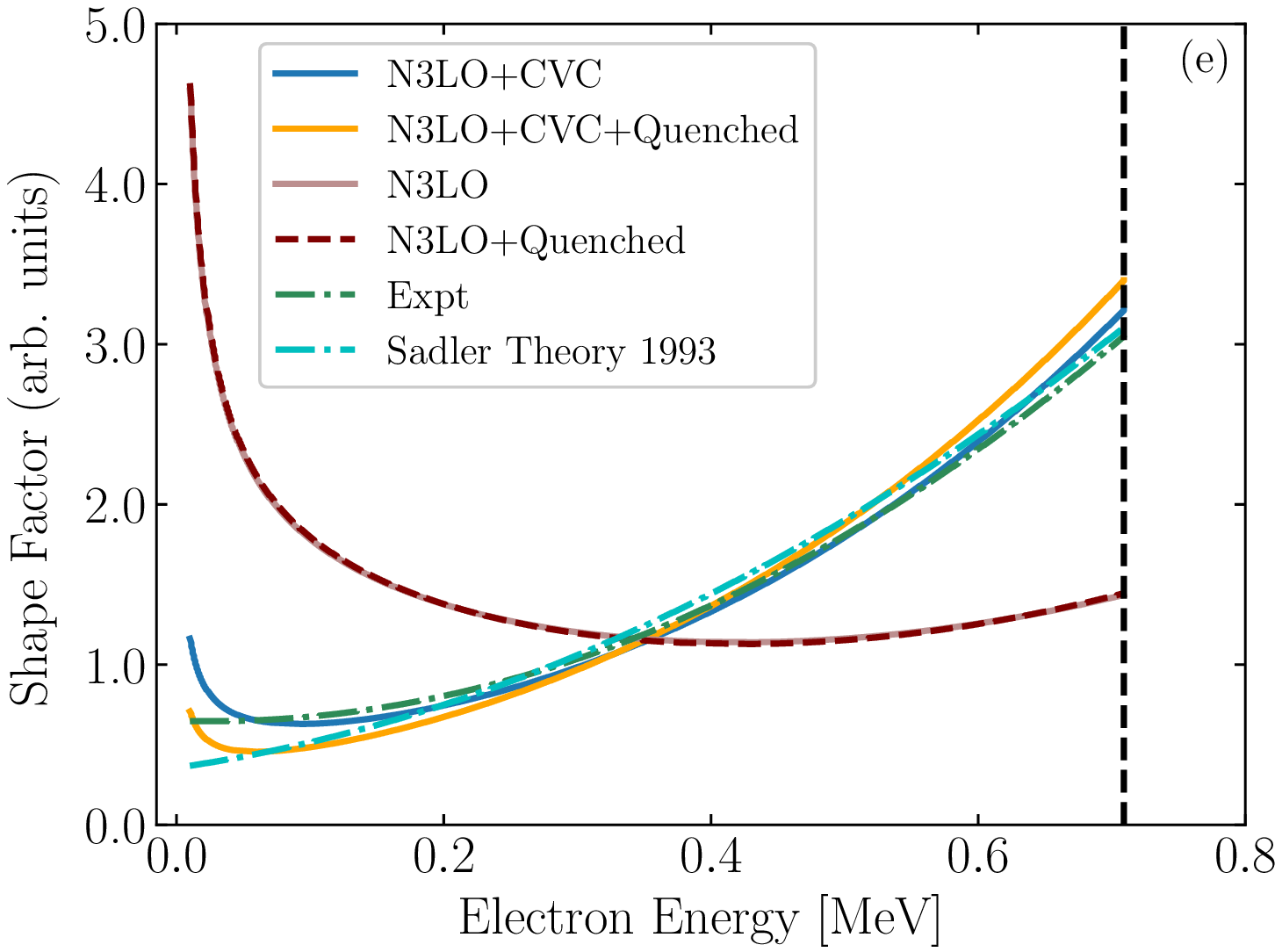}
\includegraphics[width=0.90\columnwidth]{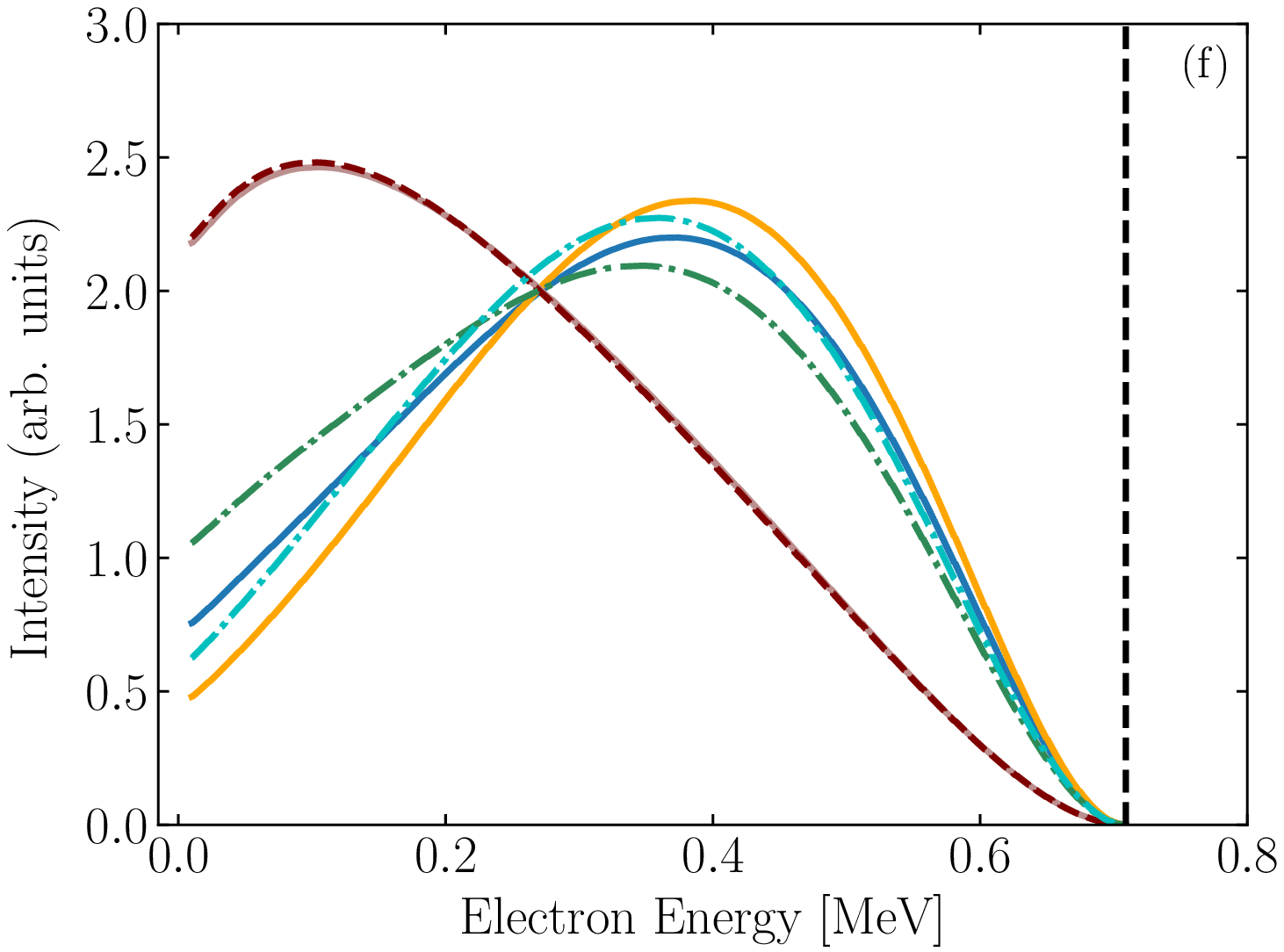}
\includegraphics[width=0.90\columnwidth]{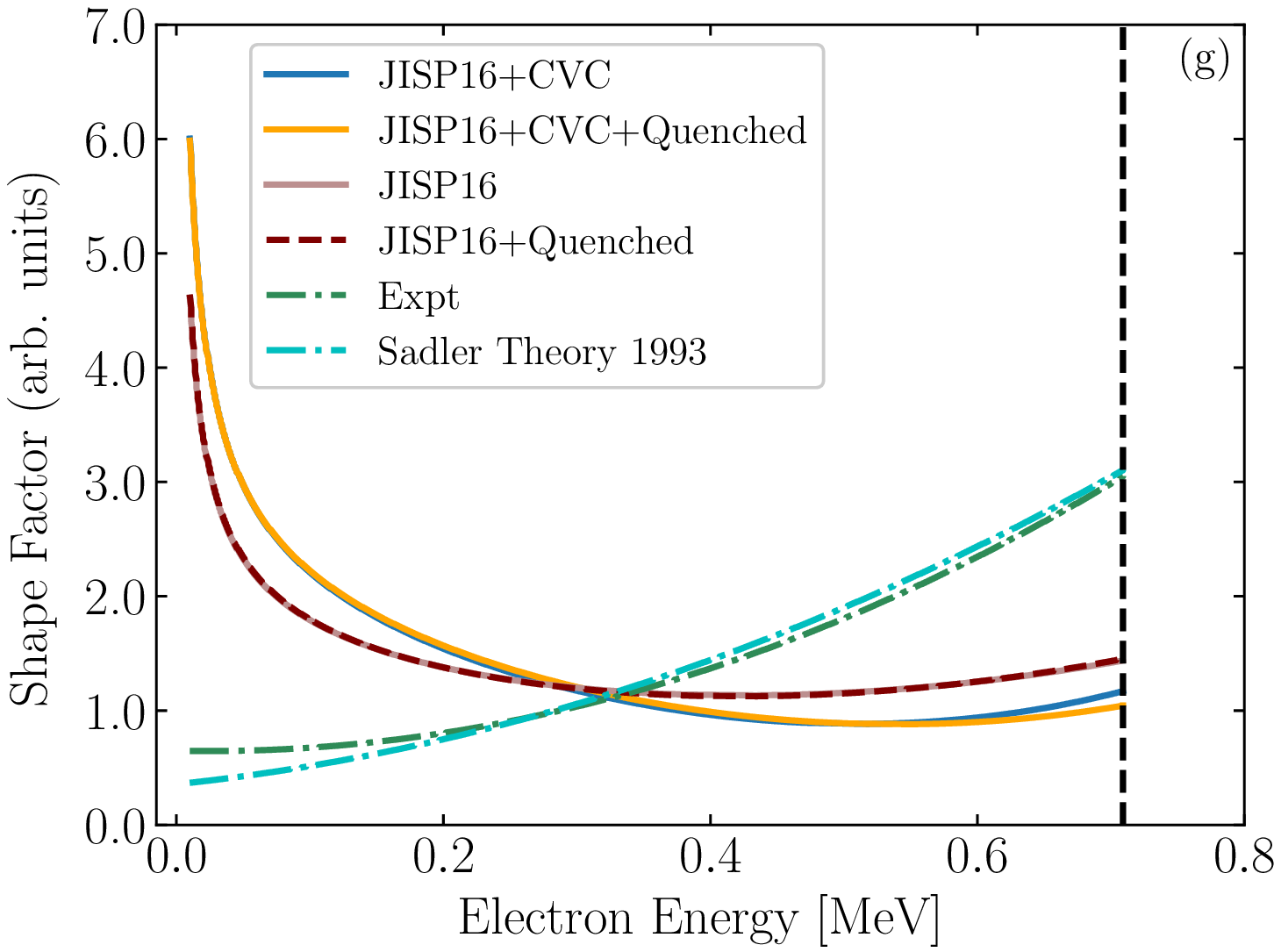}
\includegraphics[width=0.90\columnwidth]{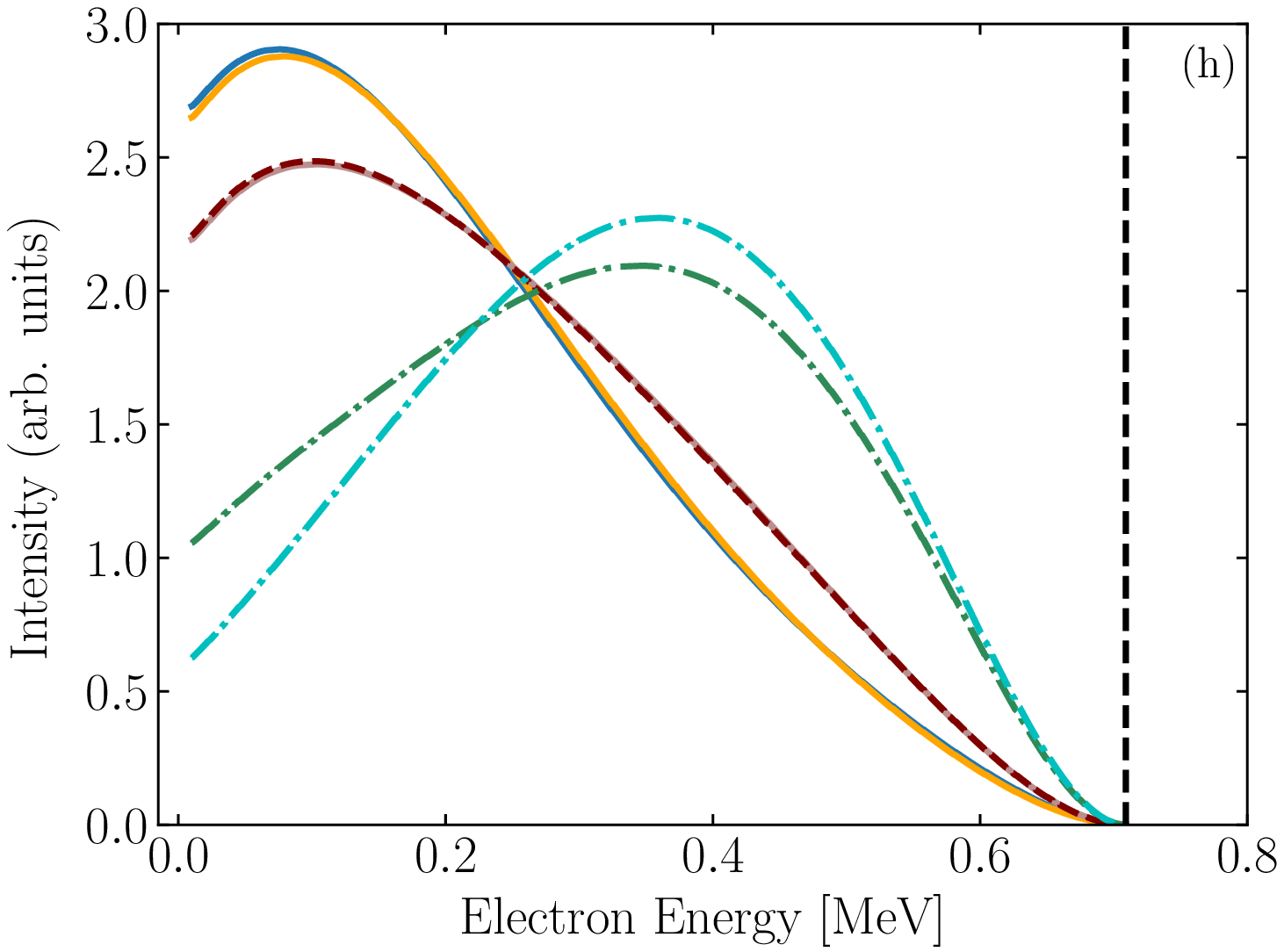}
 \caption{\label{36cl_shape} Theoretical  shape factors  (left panel) and electron spectra (right panel)  for 
 second forbidden $\beta^-$ decay of $^{36}$Cl($2^+$)$\rightarrow^{36}$Ar($0^+$) as 
 functions of electron kinetic energy for different  cases. The dashed vertical lines indicate 
 the end-point energy for forbidden ($Q_\text{forbidden}$)  decay. The area under each curve are normalized to unity.} 
\end{figure*}

\begin{figure*}
%[!ht]
\centering
\includegraphics[width=0.90\columnwidth]{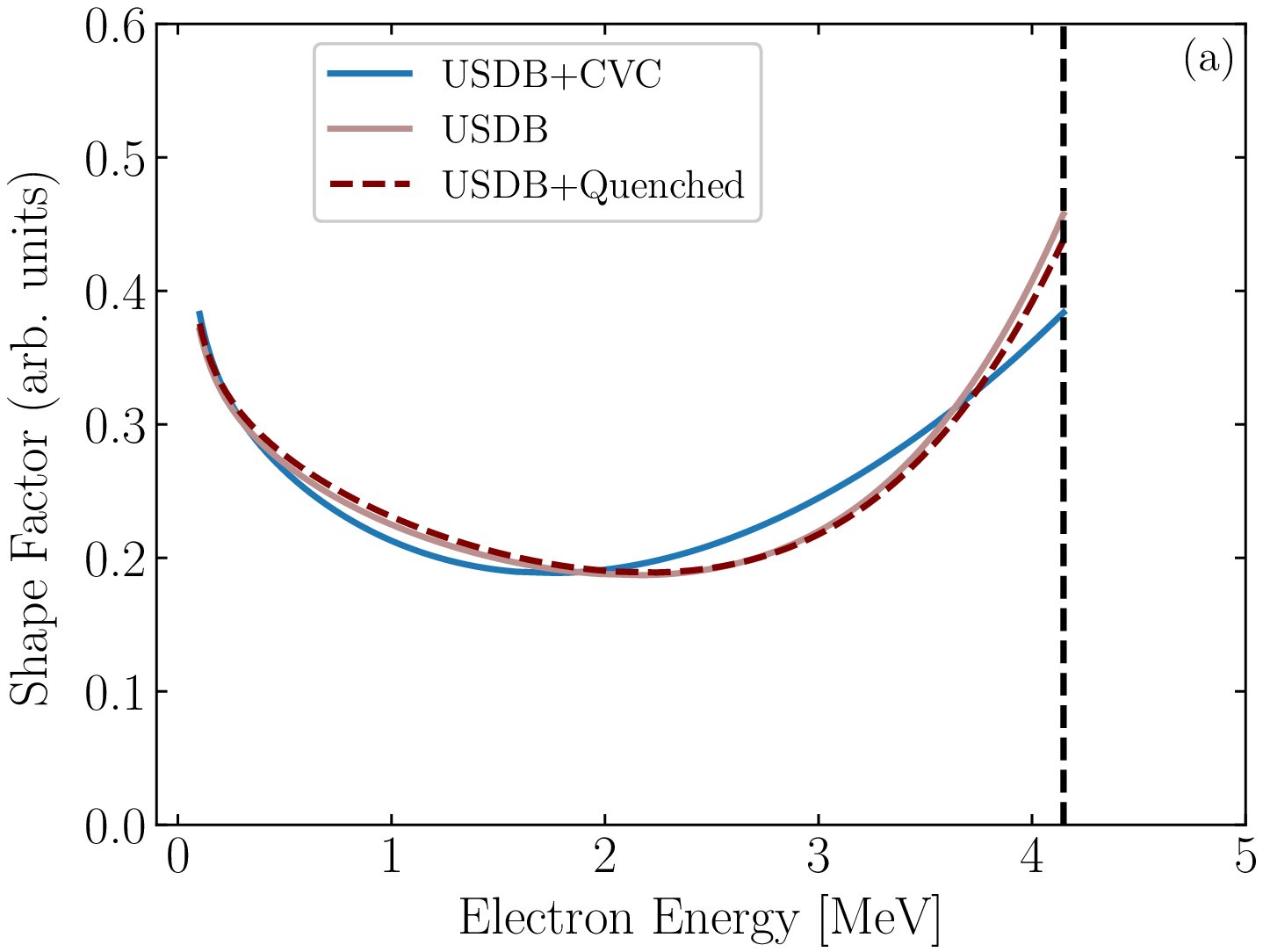}
\includegraphics[width=0.90\columnwidth]{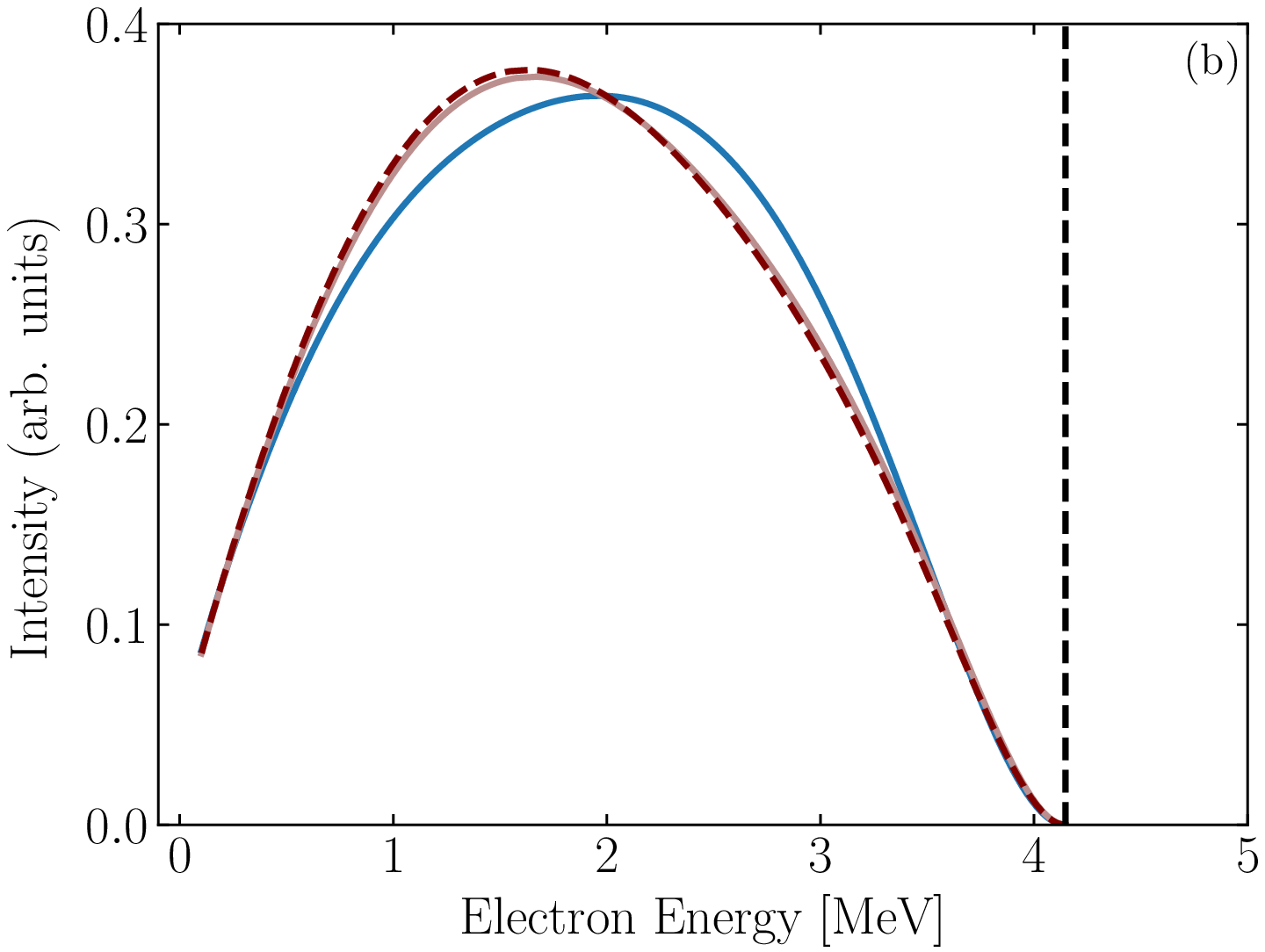}
\includegraphics[width=0.90\columnwidth]{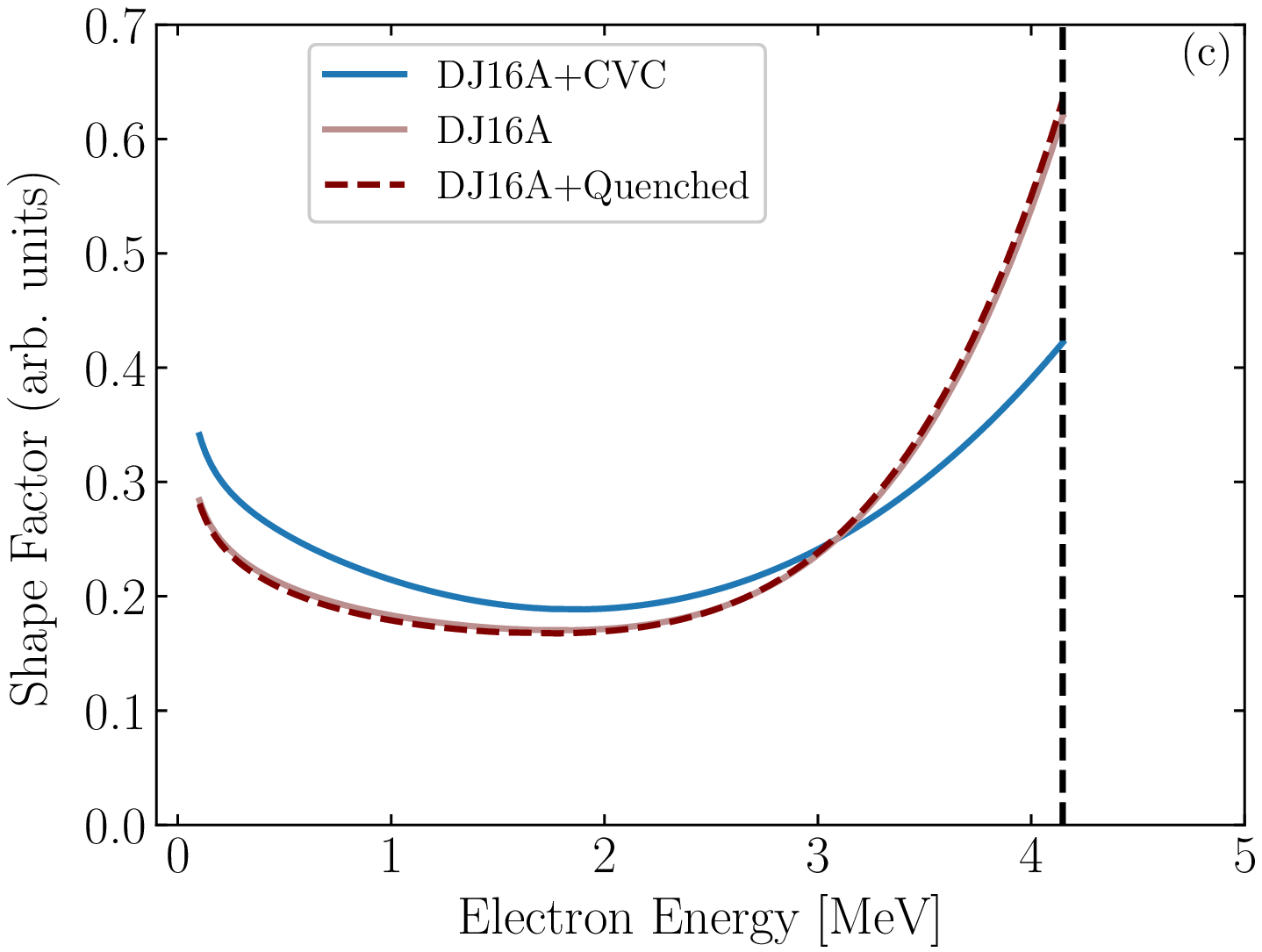}
\includegraphics[width=0.90\columnwidth]{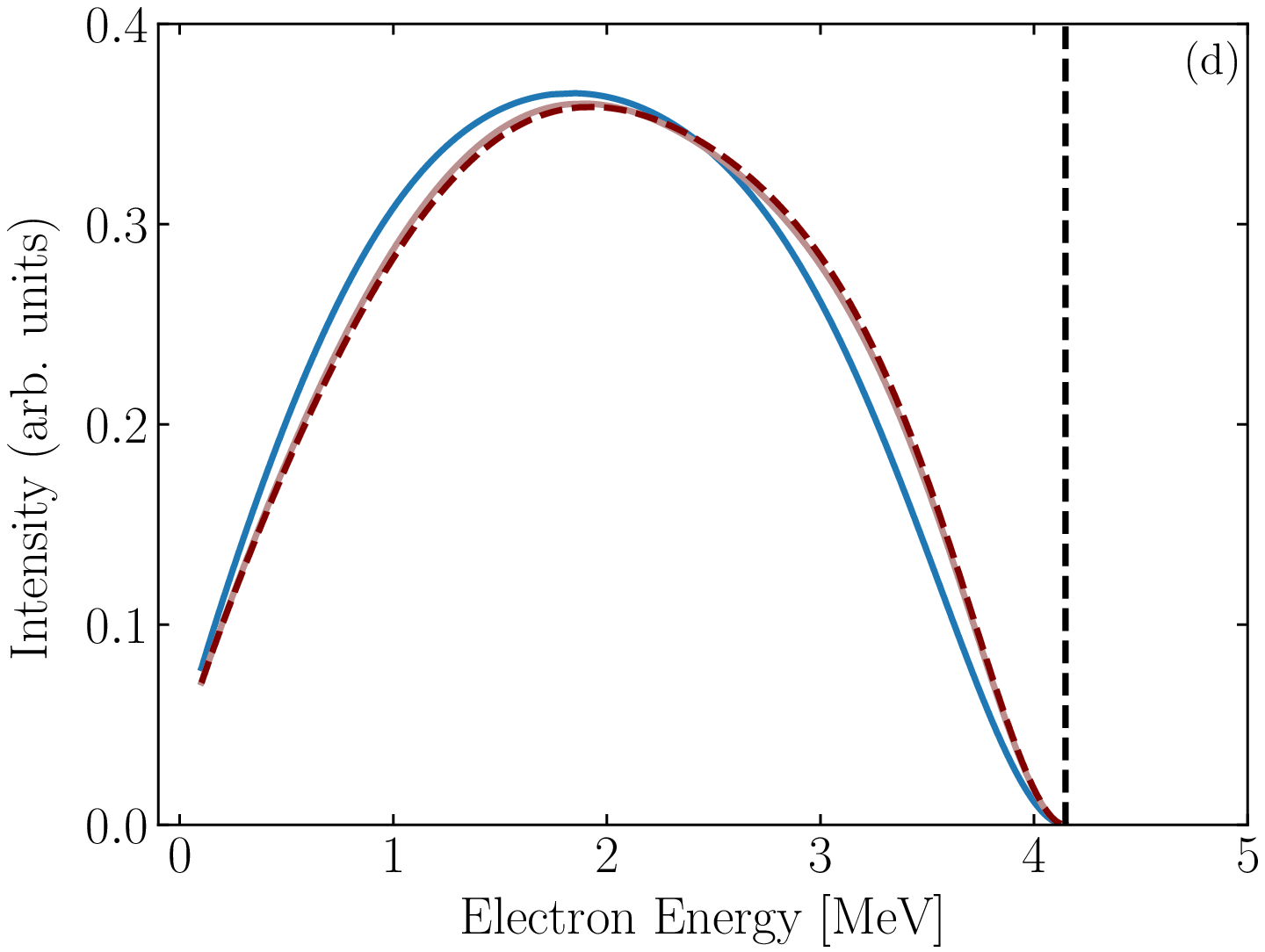}
\includegraphics[width=0.90\columnwidth]{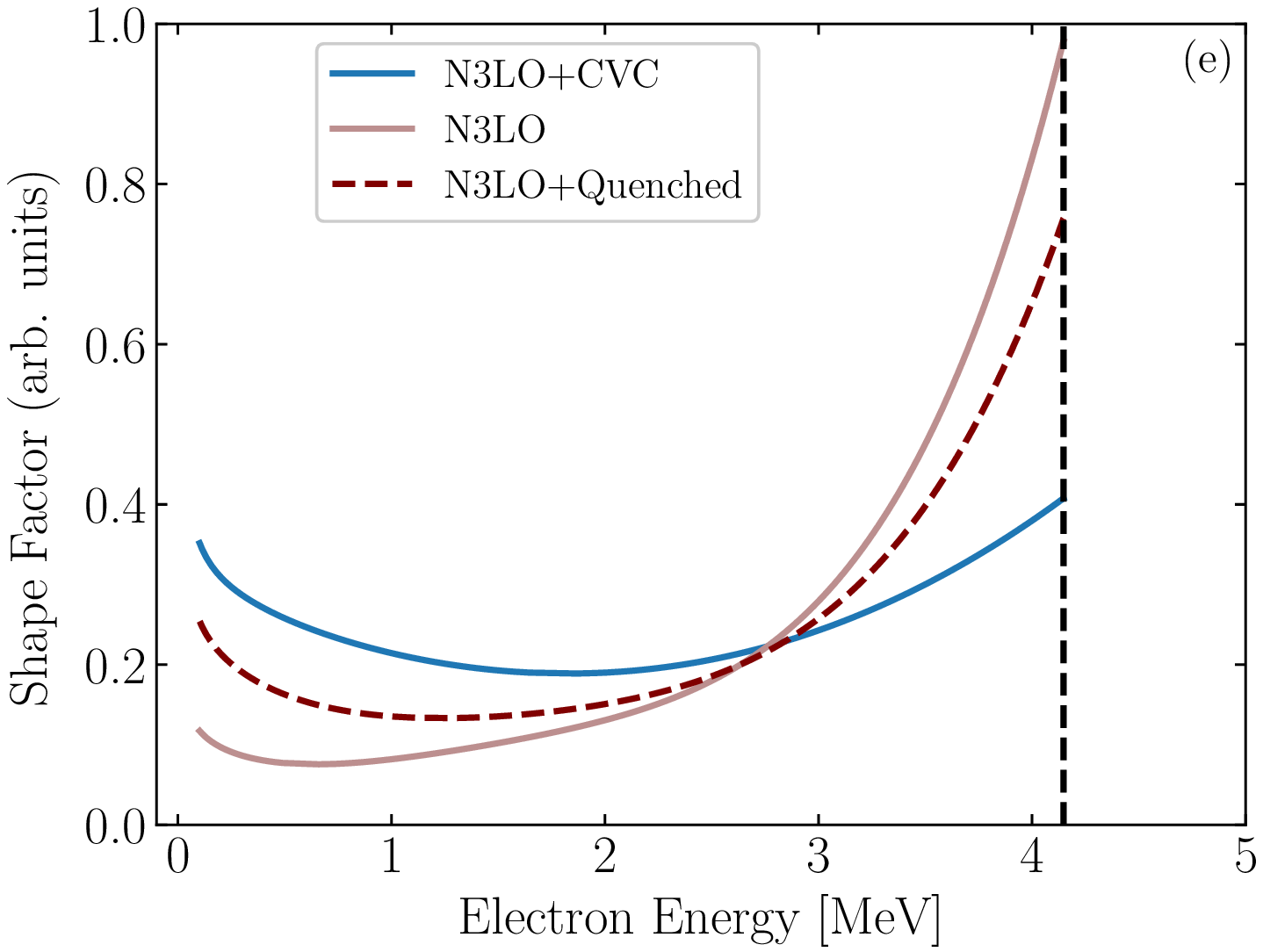}
\includegraphics[width=0.90\columnwidth]{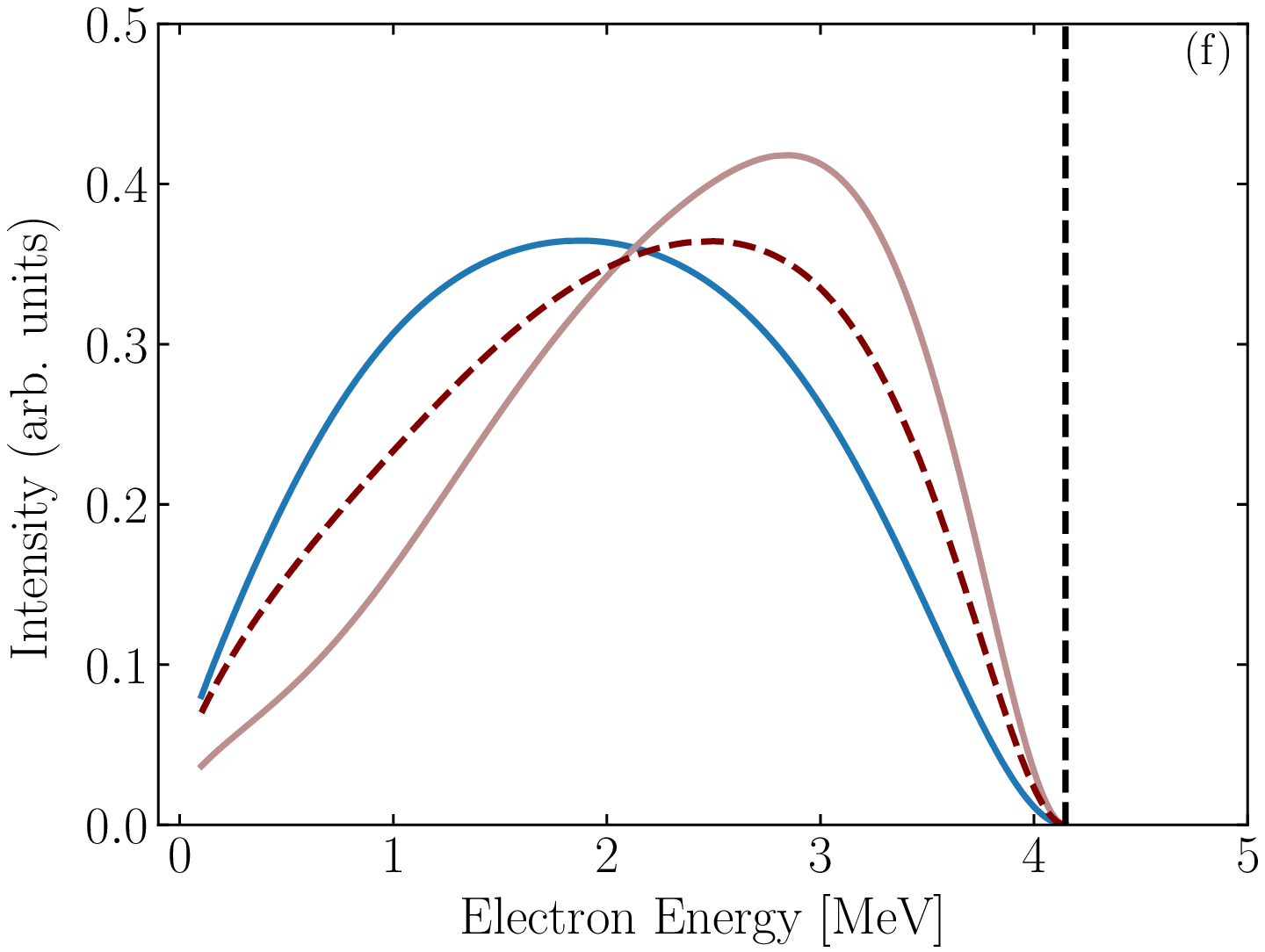}
\includegraphics[width=0.90\columnwidth]{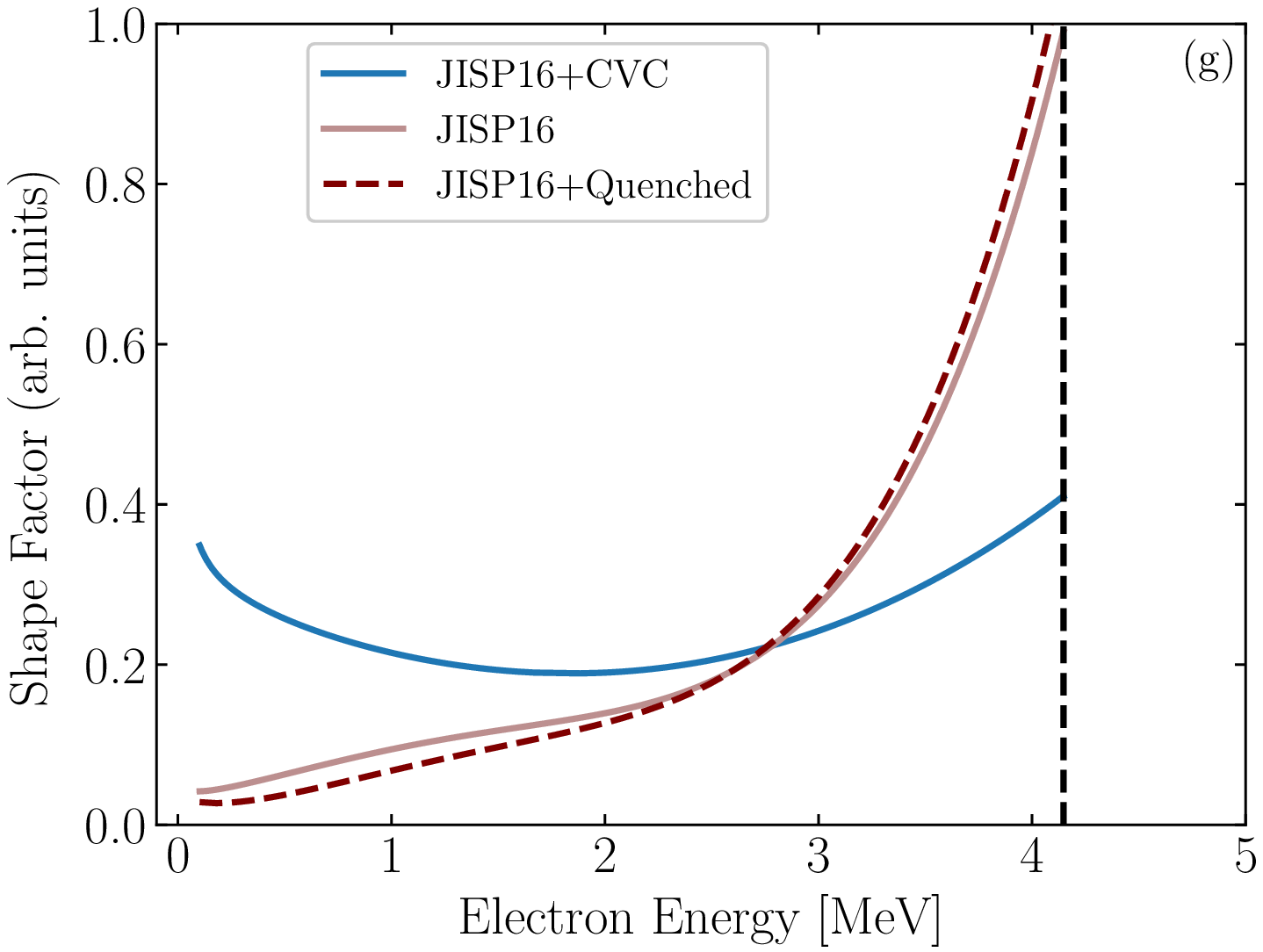}
\includegraphics[width=0.90\columnwidth]{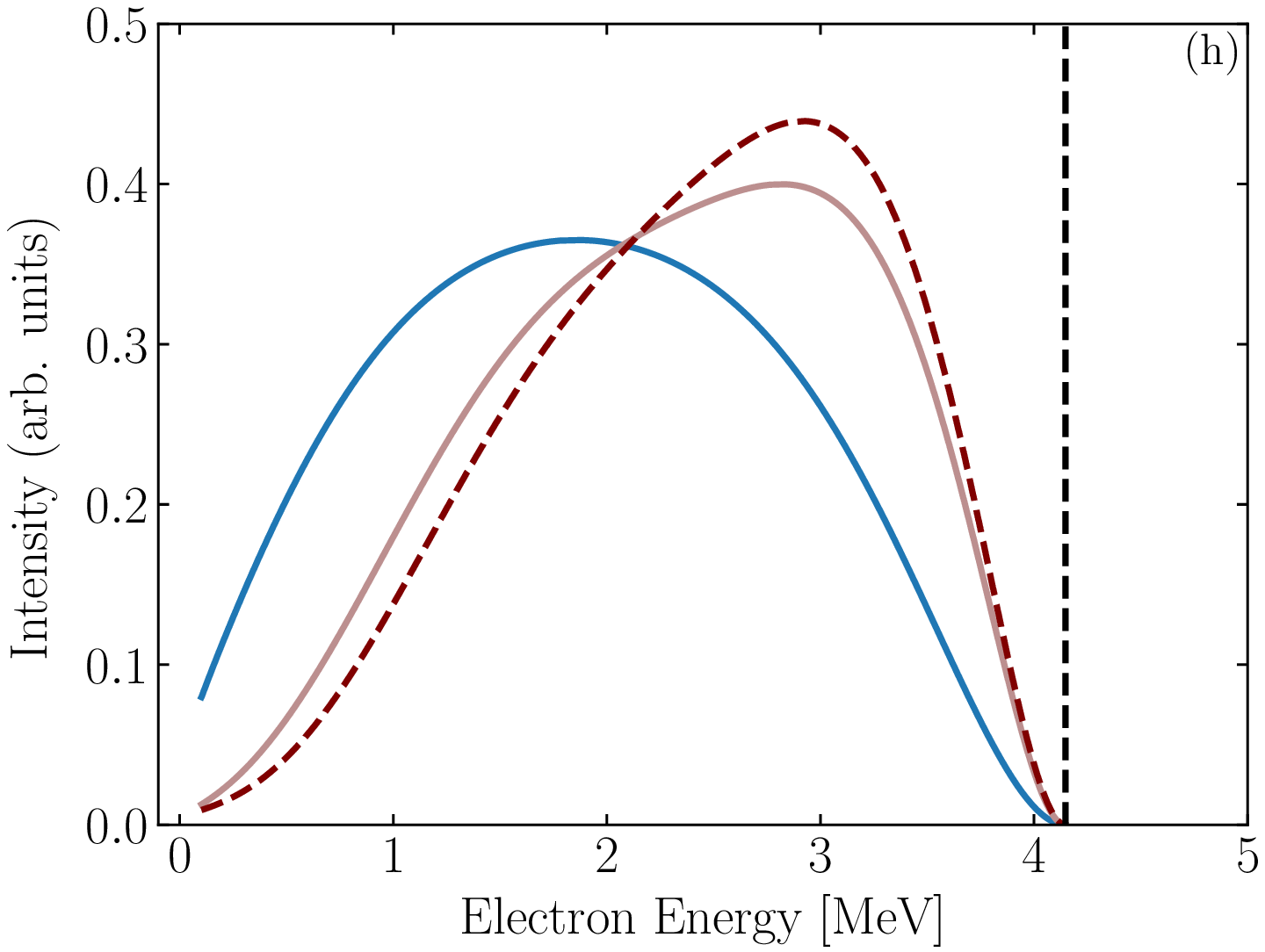}
\caption{\label{24na_shape} Theoretical  shape factors  (left panel) and electron spectra (right panel) for second order forbidden $\beta^-$ decay of $^{24}$Na($4^+$)$\rightarrow^{24}$Mg($2^+$) as functions of electron
kinetic energy for  different cases. The dashed vertical lines indicate the end-point energy for forbidden 
($Q_\text{forbidden}$) decay. The area under each curve are normalized to unity.} 
\end{figure*}

\begin{table*}
%[!ht]
\vspace{-1ex}
%\begin{small}
%\centering
\caption{\label{itc_tab} The dimensionless integrated shape factors $\tilde{C}$ for
the studied transitions,  and their decompositions to vector $\tilde{C}_V$, axial-vector $\tilde{C}_A$, and vector-axial-vector $\tilde{C}_{VA}$ parts. For the calculation of total integrated shape factor $\tilde{C}$ we have taken  $g_V=g_A=1.0$.}
\begin{ruledtabular}
\begin{tabular}{lcccc} \\

%24Na>>>>Energy_24Mg
&   \multicolumn{4}{c}{$^{24}$Na($4^+$)$\rightarrow${$^{24}$Mg($2^+$)(\text{SM})}} \\

\cline{2-5}
Interactions &$\tilde{C}_V$ &$\tilde{C}_A$ & $\tilde{C}_{VA}$ &  $\tilde{C}$ \\
\hline 
%\rule{0pt}{3ex}

USDB      &  1.3982$\times{10^{-6}}$ & 3.6829$\times{10^{-6}}$ &  1.7841$\times{10^{-6}}$ & 6.8653$\times{10^{-6}}$  \\
DJ16A     &  9.8078$\times{10^{-8}}$ & 1.6952$\times{10^{-6}}$ & -5.1005$\times{10^{-7}}$ & 1.2833$\times{10^{-6}}$  \\
N3LO      &  3.7996$\times{10^{-7}}$ & 2.2572$\times{10^{-7}}$ & -5.0839$\times{10^{-7}}$ & 9.7291$\times{10^{-8}}$   \\
JISP16    &  2.2998$\times{10^{-7}}$ & 3.7051$\times{10^{-7}}$ & -5.0759$\times{10^{-7}}$ & 9.2903$\times{10^{-8}}$   \\ 

\hline \\

&   \multicolumn{4}{c}{$^{24}$Na($4^+$)$\rightarrow${$^{24}$Mg($2^+$)(\text{SM+CVC})}} \\

\cline{2-5}

Interactions &$\tilde{C}_V$ &$\tilde{C}_A$ & $\tilde{C}_{VA}$ &  $\tilde{C}$ \\
\hline

USDB      &  8.3097$\times{10^{-5}}$ & 3.6829$\times{10^{-6}}$ & -1.3924$\times{10^{-5}}$ & 7.2856$\times{10^{-5}}$  \\
DJ16A     &  5.8298$\times{10^{-5}}$ & 1.6952$\times{10^{-6}}$ &  1.2868$\times{10^{-5}}$ & 7.2861$\times{10^{-5}}$  \\
N3LO      &  6.5792$\times{10^{-5}}$ & 2.2572$\times{10^{-7}}$ &  6.8432$\times{10^{-6}}$ & 7.2861$\times{10^{-5}}$   \\
JISP16    &  6.3828$\times{10^{-5}}$ & 3.7051$\times{10^{-7}}$ &  8.6647$\times{10^{-6}}$ & 7.2864$\times{10^{-5}}$   \\ 

\hline
\hline\\

% 36Cl>>>36Ar

&   \multicolumn{4}{c}{$^{36}$Cl($2^+$)$\rightarrow${$^{36}$Ar($0^+$)(\text{SM})}} \\

\cline{2-5}

Interactions &$\tilde{C}_V$ &$\tilde{C}_A$ & $\tilde{C}_{VA}$ &  $\tilde{C}$ \\
\hline

USDB      &  6.0691$\times{10^{-9}}$ & 3.1198$\times{10^{-10}}$ & -2.7292$\times{10^{-9}}$ & 3.6519$\times{10^{-9}}$  \\
DJ16A     &  2.1890$\times{10^{-9}}$ & 8.1048$\times{10^{-10}}$ & -2.6419$\times{10^{-9}}$ & 3.5761$\times{10^{-10}}$  \\
N3LO      &  3.8787$\times{10^{-9}}$ & 5.9273$\times{10^{-10}}$ & -3.0074$\times{10^{-9}}$ & 1.4641$\times{10^{-9}}$   \\
JISP16    &  4.4736$\times{10^{-9}}$ & 5.4657$\times{10^{-10}}$ & -3.1016$\times{10^{-9}}$ & 1.9186$\times{10^{-9}}$   \\ 

\hline \\

&   \multicolumn{4}{c}{$^{36}$Cl($2^+$)$\rightarrow${$^{36}$Ar($0^+$)(\text{SM+CVC})}} \\

\cline{2-5}

Interactions &$\tilde{C}_V$ &$\tilde{C}_A$ & $\tilde{C}_{VA}$ &  $\tilde{C}$ \\
\hline

USDB      &  4.0126$\times{10^{-10}}$ & 3.1198$\times{10^{-10}}$ & -7.7968$\times{10^{-11}}$ & 6.3528$\times{10^{-10}}$  \\
DJ16A     &  1.4791$\times{10^{-10}}$ & 8.1048$\times{10^{-10}}$ & -3.2311$\times{10^{-10}}$ & 6.3528$\times{10^{-10}}$  \\
N3LO      &  2.5097$\times{10^{-10}}$ & 5.9273$\times{10^{-10}}$ & -2.0843$\times{10^{-10}}$ & 6.3527$\times{10^{-10}}$   \\
JISP16    &  2.2999$\times{10^{-9}}$ & 5.4657$\times{10^{-10}}$ & -2.2112$\times{10^{-9}}$ & 6.3528$\times{10^{-10}}$   \\

\end{tabular}
\end{ruledtabular}
%\end{small}
\end{table*}

\subsection{Shape Factors and Electron Spectra} \label{ele_spec}

In Fig. \ref{36cl_shape} and \ref{24na_shape}, we have shown the shape factors (left panel) and $\beta$ spectra (right panel) of the second forbidden
nonunique $\beta^-$ decays of  $^{36}$Cl and $^{24}$Na.  The second-forbidden nonunique beta decay of $^{36}$Cl is   predicted  with strong branching ratio 98.1\%, while that of $^{24}$Na is predicted with  a weak branching ratio less than 1\%. These figures represent the shape factor of Eq. (\ref{eq2}) and $\beta$ spectrum corresponding to the integrand of Eq. (\ref{tc}) as a function of electron kinetic energy for different microscopic  and USDB effective  interactions. For all these  calculations  of second-forbidden nonunique beta decay of $^{24}$Na and $^{36}$Cl, we have used the experimentally measured $Q$ value $4147$ KeV and $709.547$ KeV, respectively.
We have calculated the shape factor by including only the leading-order terms, and  the value of vector coupling constant $g_{V}=1.00$ were adopted by CVC hypothesis. We presented in figures the purely theoretical results from the shell model interactions, labeled ``name of interactions,''  and those constrained from experimental information labeled  ``name of interactions and CVC theory''  with quenched ($g_{A}=1.00$) or bare ($g_A=1.27$) cases. The areas under both the theoretical and experimental curves are normalized to unity.

For the shape factor and $\beta$ spectrum of $^{36}$Cl, we  have done  a comparison with the available experimental data due to Rotzinger $et ~ al$ \cite{Rotzinger2008} and  with the theoretical  results of Sadler $et ~  al$ \cite{Sadler1993}.
%The experimentally measured $Q$ value  $709.547$ KeV \cite{nndc}  is used in our study for  $^{36}$Cl. 
In the case of $^{36}$Cl, the shape factor calculated with the matrix element ``$^V\mathcal{M}_{211}^{(0)}=0$'' 
yields a poor agreement in comparison to the experimental shape factor. 
After constraining this matrix element with the experimental half-life, the shape factor and electron spectra are consistent with  the  experimental data. The electron spectra from ``DJ16A+CVC+Quenched'' are perfectly  matched  with the experimental electron  spectra.  This means that the shape factor and electron spectra strongly depend on this matrix element $^V\mathcal{M}_{211}^{(0)}$.    
But in the  case of JISP16 interaction, we  have  not obtained   a  good number of this matrix elements from  the  experimental half-life method.  We have obtained the value of the matrix element    $^V\mathcal{M}_{211}^{(0)} = -0.007451\pm0.0009$  for JISP16 interaction,  it  is too small as compared to other interactions.

In Fig. \ref{24na_shape}, we have presented the shape factor and $\beta$ spectrum of $^{24}$Na from purely shell model calculation with quenched and unquenched cases. In the pure shell-model calculations, the shape-factor and $\beta$-spectrum curves  depend  strongly on the quenching value of $g_A$.  After  CVC constraining the matrix element ``$^V\mathcal{M}_{211}^{(0)}$'', we find that the shape factor and $\beta$ spectrum are 
independent of the value of $g_A$.  So, we have presented curve for ``SM+CVC'' only for bare $g_A$ value.
 %For all these calculation of second forbidden beta decay $^{24}$Na, we have used the experimentally measured $Q$ value 4147 KeV. 
 For the comparison, there are no experimental data available for shape factor and electron spectra corresponding to the second-forbidden nonunique $\beta^-$  decay of $^{24}$Na.  Thus, our theoretical results might be quite useful to compare with a future experimental measurement.

\subsection{Decomposition of the integrated shape factor} \label{shape_d}

In Table \ref{itc_tab}, we present the integrated shape factor $\tilde{C}$ and its decomposition to vector $\tilde{C}_V$, axial-vector $\tilde{C}_A$, and mixed vector-axial-vector $\tilde{C}_{VA}$ components, for the   involved transitions using different effective interactions. Hence, we have calculated the value of $\tilde{C}$ and its components with purely shell model labeled ``SM'' and after putting  constrained to the  matrix element $^V\mathcal{M}_{211}^{(0)}$ from experimental information labeled, ``SM+CVC''. For all the studied decays transition, the sign of vector $\tilde{C}_V$ and axial-vector $\tilde{C}_A$ components is positive from  ``SM'' and ``SM+CVC'', but the sign of mixed axial-vector  $\tilde{C}_{VA}$ component varies. 
From the pure ``SM'' for $^{24}$Na, the axial-vector component $\tilde{C}_A$  is  dominant in the USDB and DJ16A interactions. For N3LO and JISP16 interactions, the mixed component  $\tilde{C}_{VA}$ is roughly the sum of vector and axial-vector components and negative in sign. 
In ``SM+CVC'', the vector component $\tilde{C}_V$ is dominant for all interactions. The mixed component $\tilde{C}_{VA}$ is negative for USDB, while positive for other interactions. 
In case of $^{36}$Cl, the vector component $\tilde{C}_V$ is dominant for all the interactions in the case of pure ``SM''. After  applying  CVC theory, the vector part is dominant only in USDB and JISP16 interactions and for the other two interactions the axial-vector part  is large as  compared to other two components.   
The sign of  the mixed components $\tilde{C}_{VA}$ are negative in both cases ``SM'' and ``SM+CVC'' for all interactions.  

%%%%%%%%%%%%%%%%%%%%%%%%%%%%%%%%%%%%%%%%%%%%%%%%%%%%%%%%%%%%%%%%%%%%%%

\section{Conclusions}\label{Conclusions}

 In this article we have calculated log$ft$ values, shape factors and electron spectra for the second-forbidden nonunique $\beta^-$ transitions of $^{24}$Na($4^+$)$\rightarrow${$^{24}$Mg($2^+$)} and $^{36}$Cl($2^+$)$\rightarrow${$^{36}$Ar($0^+$)}   using the three  microscopic effective interactions (DJ16A, N3LO, and JISP16) obtained from the NCSM wave functions via the OLS transformation.  Also, for the comparison, we have used the more popular phenomenological effective USDB interaction.
  
 The low-lying energy spectra of the involved mother and daughter nuclei in $\beta^-$-decay corresponding to different $ab~initio$  and phenomenological effective interactions are compared with the available experimental data.
  The obtained wave functions have been used for further calculations.
To calculate the log$ft$ values, shape factors and electron spectra, we have constrained the relativistic matrix element $^V\mathcal{M}_{211}^{(0)}$ in  the $sd$ model space by experimental information.  This matrix element  plays an important role in the shape factor and electron spectra. The calculated log$ft$ values are compared with experimental data. In
the  case of JISP16 interaction, we  could not obtain a  proper value  of this matrix  element. 
In our calculation, we have used two different  values of $g_A$, either  the bare value of $g_A=1.27$ or the quenched value of $g_A=1.00$. 
For the allowed beta decay of $^{24}$Na, the log$ft$ values are in reasonable agreement with the experimental data. In case of second-forbidden non unique beta decay, we have calculated log$ft$ values corresponding to $g_A=1.27$ and compared with the experimental data. 
Before CVC theory the electron spectra of $^{24}$Na depend significantly on the effective value of $g_A$, while after CVC  it  has  become  independent. In  the  case of $^{36}$Cl, the dependency of electron spectra on $g_A$  is  opposite from the case of $^{24}$Na for USDB, N3LO, and JISP16 interactions, but in case of DJ16A interaction the electron spectra strongly depend on $g_A$ before and after CVC theory. 
In case of $^{36}$Cl, the experimental  data  are  available for shape factors and electron spectra. So we have  compared our  theoretical results with the  experimental data to check the role of matrix element $^V\mathcal{M}_{211}^{(0)}$. But in the  case of $^{24}$Na, there are no experimental data available for shape factor and electron spectra. Thus,  our calculated results could be quite  useful when compared with future experimental data. 
Also, we  have  decomposed the integrated shape function $\tilde{C}$ in to vector $\tilde{C}_V$, axial-vector $\tilde{C}_A$, and vector-axial-vector $\tilde{C}_{VA}$ components  to  see the individual effect of these components.

%\newpage
\section{Acknowledgments}\label{Acknowledgments}
AK would like to thanks for financial support from the Ministry of Human Resource Development (MHRD), Government of India, for his thesis work. PCS acknowledges a research grant from SERB (India), CRG/2019/000556.   
 We would like to thank X. Mougeot for providing the experimental shape factor data of $^{36}$Cl. 

\bibliographystyle{utphys}
  \bibliography{references}

%\bibliography{bibliography}
%\bibliography{xnew_incoherent.bbl}

\end{document}